\documentclass[preprint]{elsarticle}

\usepackage{amsmath}
\usepackage{amssymb}
\usepackage{graphicx}
\usepackage{color} 
\usepackage{todonotes}
\usepackage{hyperref}
\usepackage{movie15}

\usepackage{latexsym}
\usepackage[english]{babel}
\usepackage[utf8]{inputenc}
\usepackage[squaren]{SIunits}
\usepackage{color} 
\usepackage{todonotes}

\newcommand*{\vct}[1]{\ensuremath{\boldsymbol{#1}}}
\newcommand*{\bigo}[1]{\ensuremath{\mathcal{O}\left(#1\right)}}

\newcommand*{\smallo}[1]{\ensuremath{o\left(#1\right)}}

\newcommand*{\zlimpm}{\ensuremath{\lim_{z\to\pm\infty}}}

\newcommand*{\grad}{\boldsymbol\nabla}

\newcommand*{\gradG}{\boldsymbol\nabla_{\Gamma}}

\newcommand*{\laplG}{\boldsymbol\Delta_{\Gamma}}
\newcommand*{\lapl}{\boldsymbol\Delta}

\newcommand*{\ndot}{\vct n\cdot}

\begin{document}

\begin{frontmatter}
\title{Diffuse interface models of locally inextensible vesicles in a viscous fluid}

\author[label1]{Sebastian Aland}
\ead{sebastian.aland@tu-dresden.de}
\author[label1]{Sabine Egerer}
\ead{sabine.egerer@tu-dresden.de}
\author[label2]{John Lowengrub}
\ead{lowengrb@math.uci.edu}
\author[label1]{Axel Voigt}
\ead{axel.voigt@tu-dresden.de}
\address[label1]{Institut f\"ur wissenschaftliches Rechnen, TU Dresden, 01062 Dresden, Germany}
\address[label2]{Department of Mathematics, and 
Department of Biomedical Engineering, UC Irvine, Irvine, CA 92697, USA}

\begin{abstract}
We present a new diffuse interface model for the dynamics of inextensible vesicles in a viscous fluid. A new feature of this work is the implementation of the local inextensibility condition in the diffuse interface context. Local inextensibility is enforced by using a local Lagrange multiplier, which provides the necessary tension force at the interface. To solve for the local Lagrange multiplier, we introduce a new equation whose solution essentially provides a harmonic extension of the local Lagrange multiplier off the interface while maintaining the local inextensibility constraint near the interface. To make the method more robust, we develop a local relaxation scheme that dynamically corrects local stretching/compression errors thereby preventing their accumulation. Asymptotic analysis is presented that shows that our new system converges to a relaxed version of the inextensible sharp interface model. This is also verified numerically. Although the model does not depend on dimension, we present numerical simulations only in 2D. To solve the 2D equations numerically, we develop an efficient algorithm combining an operator splitting approach with adaptive finite elements where the Navier-Stokes equations are implicitly coupled to the diffuse interface inextensibility equation. Numerical simulations of a single vesicle in a shear flow at different Reynolds numbers demonstrate that errors in enforcing local inextensibility may accumulate and lead to large differences in the dynamics in the tumbling regime and differences in the inclination angle of vesicles in the tank-treading regime. The local relaxation algorithm  is shown to effectively prevent this accumulation by driving the system back to its equilibrium state when errors in local inextensibility arise.

\end{abstract}


\end{frontmatter}

\section{Introduction}
\label{introduction}

Vesicles are fluid-filled sacs bounded by a closed lipid bilayer membrane. Vesicles play a critical role in intracellular transport  of molecules and proteins \cite{Alberts}. Vesicles have been used as drug delivery vehicles \cite{Sofou07}, microreactors \cite{Fischer02} and as models of more complex biostructures such as red blood cells (RBCs) \cite{Siefert97}. RBCs and vesicles are known to undergo complex motions and shape changes under applied flows (e.g., \cite{Fischeretal_Science_1978,Kantsleretal_PRL_2006,AbkarianViallat2008,Dechamps2009,Brown2011,VlahovskaEtAl2013,LiVlahovskaKarniadakis2013}) and transitions from stationary shapes (tank-treading, TT) to trembling (TR) to tumbling (TB) have been observed as a function of flow conditions and membrane characteristics. RBCs resist shear deformation due to the presence of a membrane cytoskeleton and also resist bending and area dilatation (e.g., \cite{AbkarianStone2008,WanEtAl2011,VlahovskaEtAl2013}), while the lipid bilayer membranes in vesicles  are liquid-like, resist bending and are largely inextensible (e.g., \cite{Lipowsky1991,Siefert97}). In this paper, we focus on the dynamics of homogeneous vesicles, although our results apply more generally to the case in which there may be several lipid components on the membrane that can induce the formation of rafts. 

Most experimental results on vesicles are performed in the low Reynolds number regime, see e.g. \cite{Kantsleretal_PRL_2006,Maderetal_EPJE_2006,Dechamps2009}. Under these conditions inertia effects can be neglected and the Stokes limit considered, which allows the development of small-deformation perturbation theories \cite{Misbah_PRL_2007,Dankeretal_PRE_2007,Lebedevetal_PRL_2007,Noguchietal_PRL_2007,
Vlahovskaetal_PRE_2007,Kaourietal_PRE_2009,Messlingeretal_PRE_2009}, which all qualitatively predict the experimentally observed TT and TB motion. Various numerical approaches have also been considered in the Stokes limit to analyze TT and TB motions, e.g. \cite{Kelleretal_JFM_1982,Krausetal_PRL_1996,
Bibenetal_PRE_2003,Beaucourtetal_PRE_2004,Bibenetal_PRE_2005,Veerapanemietal_JCP_2009,
RahimianEtAl2010,Ghigliottietal_JFM_2010,Sohnetal_JCP_2010,Kimetal_JCP_2010,Zhaoetal_JFM_2011,VeerapaneniEtAl2011,PecoEtAl2013}.  Except for \cite{Kelleretal_JFM_1982} in which the vesicle shape was assumed to be a fixed ellipsoid, all other models are of Helfrich type and consider a membrane free energy
\begin{eqnarray}
{\cal{E}} = \int_\Gamma \frac{1}{2}b_N (H - H_0)^2 \; d\Gamma+\int_\Gamma b_G K\:d\Gamma \label{eq:Helfrich_sharp}
\end{eqnarray}
with membrane $\Gamma(t)$, mean curvature $H$, spontaneous curvature $H_0$, normal bending rigidity $b_N$, Gaussian bending rigidity $b_G$ and Gaussian curvature $K$. We focus on the case in which the vesicle is homogeneous and its topology does not change. Then $b_N$, $H_0$ and $b_G$ may be assumed to be constant and the Gaussian bending energy only contributes a constant and can therefore be neglected. Lagrange multipliers are used to enforce the inextensibility constraint, which can be considered as a global constraint to enforce a constant area of the membrane, but allowing for local variations, or as a stronger local constraint. The jump condition for the fluid stress tensor $\mathbf{S} = -p \mathbf{I} + \nu \mathbf{D}$, where $p$ is the pressure, $\nu$ is the viscosity, and $\mathbf{D}$ is the rate of deformation tensor $\mathbf{D} =\left( \nabla \mathbf{v} + (\nabla \mathbf{v})^T\right)/2$, with velocity $\mathbf{v}$, along the membrane then reads
\begin{align}
\!\!\! [\mathbf{S} \cdot \mathbf{n}]_\Gamma \!\!&=\!\! \frac{\delta {\cal{E}}}{\delta \Gamma} &\qquad \mbox{unconstrained,} \\
\!\!\! [\mathbf{S} \cdot \mathbf{n}]_\Gamma \!\!&=\!\! \frac{\delta {\cal{E}}}{\delta \Gamma} + \lambda_{global} H \mathbf{n} &\qquad \mbox{global area constraint,} \\
\!\!\! [\mathbf{S} \cdot \mathbf{n}]_\Gamma \!\!&=\!\! \frac{\delta {\cal{E}}}{\delta \Gamma} + \lambda_{local} H \mathbf{n} + \nabla_\Gamma \lambda_{local} &\qquad \mbox{local inextensibility constraint},
\label{sharp local constraint}
\end{align}
where $[f]_\Gamma=f_{outer}-f_{inner}$,  $\mathbf{n}$ is the normal pointing out of the vesicle,
and $\nabla_\Gamma$ is the surface gradient $\nabla_\Gamma = \mathbf{P} \nabla$, with the projection operator $\mathbf{P} = \mathbf{I} - \mathbf{n} \otimes \mathbf{n}$. The Lagrange multipliers are functionals of the fluid velocity $\mathbf{v}$ and are obtained by requiring 
\begin{eqnarray*}
\frac{d}{dt} \int_\Gamma \; d\Gamma = \int_\Gamma H \mathbf{v} \cdot \mathbf{n} \; d \Gamma \!\!&=&\!\! 0, \qquad  \mbox{global area constraint}, \\
\nabla_\Gamma \cdot \mathbf{v} \!\!&=&\!\! 0, \qquad \mbox{local inextensibility constraint}. 
\end{eqnarray*}
We remark that locally inextensible vesicles also conserve the global surface area. The jump condition for the velocity reads in all cases
\begin{eqnarray}
\; [\mathbf{v}]_\Gamma \!=\! 0.
\end{eqnarray} 
Due to the linearity of the Stokes problem, efficient  algorithms can be derived to solve the coupled fluid-structure flow problem, e.g.  \cite{Bibenetal_PRE_2003,Sohnetal_JCP_2010,Veerapanemietal_JCP_2009,Zhaoetal_JFM_2011,VeerapaneniEtAl2011}. When inertial forces are considered, the development of efficient algorithms remains a significant challenge.

Inertial effects can become important in a variety of biophysical applications. Flowing vesicles/RBCs in larger blood vessels such as arterioles and arteries may experience Reynolds numbers of order unity or higher, especially if the vessels are constricted due to diseases such as thrombosis, e.g. \cite{VennemannEtAl2007,BarkKu2010}. Large Reynolds numbers may also be found in biomedical devices such as ventricular assist devices, e.g.,  \cite{FraserEtAl2011}. Motivated by these applications inertia effects are considered in \cite{Laadharietal_PF_2012,Salacetal_JFM_2012,MaitreEtAl2012,KimLai2012,DoyeuxEtAl2013,LuoEtAl2013}, which  found that the classical TB behavior of highly viscous vesicles is no longer observed at moderate Reynolds numbers.



%
%
%

The Navier-Stokes equations inside and outside the vesicle read
\begin{eqnarray}
\rho (\partial_t \mathbf{v} + \mathbf{v} \cdot \nabla \mathbf{v}) - \nabla \cdot \mathbf{S} \!\!&=&\!\!0 \\
\nabla \cdot \mathbf{v} \!\!&=&\!\! 0
\end{eqnarray}
with density $\rho = \rho_{1,2}$ and stress tensor $\mathbf{S} = \mathbf{S}_{1,2} = - p \mathbf{I} + \nu_{1,2} \mathbf{D}$. The global area constraint, which can be treated explicitly, has been used  by \cite{Bonitoetal_MMNP_2011} within a front tracking method, by \cite{Duetal_DCDS_2007,DuEtAl,Marthetal_JMB_2013,Hausseretal_IJBMBS_2013} within a phase field method, and was also considered in  \cite{Salacetal_JCP_2011} within a level-set approach.


The local inextensibility constraint is more delicate and leads to additional nonlinearities. This has been  considered within a level set approach in \cite{Salacetal_JCP_2011,Salacetal_JFM_2012,Laadharietal_PF_2012,DoyeuxEtAl2013}, immersed boundary methods \cite{Kimetal_JCP_2010,KimLai2012} and phase field methods \cite{Bibenetal_PRE_2003,Bibenetal_PRE_2005,MaitreEtAl2012,Laadharietal_PF_2012}. Capsule-like models have also been considered using strain-energy functions that penalize local stretching, e.g. \cite{LuoEtAl2013,CordascoBagchi2013}.

In \cite{Salacetal_JCP_2011,Salacetal_JFM_2012} the system is rewritten as a single-fluid model by considering the jump conditions for the fluid stress tensor as a body-force term with a delta-function $\delta_\Gamma$ to localize the force at the membrane. An iterative multi-step projection method is used to ensure first the incompressibility of the fluid and second to determine the Lagrange multiplier. However, the projection step to determine the Lagrange multiplier is fully explicit and does not preserve the incompressibility of the fluid. The approach also assumes that the level set function is a signed distance function, and thus requires redistancing, and that the inextensibility constraint holds in a computational domain near the interface, which can influence the velocity field in the bulk fluid phases. In \cite{Laadharietal_PF_2012,Laadharietal2013}, a saddle-point approach was used to solve the level-set formulation of the system using adaptive finite elements. An implicit time-stepping algorithm was proposed where the fluid equations and the level-set equations were solved iteratively at each time step. Additional Lagrange multipliers were introduced into the level-set equation to enhance volume and surface area conservation. Indeed, without these additional Lagrange multipliers, the volume and surface area errors increase rapidly leading to inaccuracy of the method. The additional Lagrange multipliers, however, do not introduce additional forces in the fluid, which is questionable physically. A similar approach is used in \cite{DoyeuxEtAl2013} although they did not use adaptive local refinement and did not consider the additional Lagrange multipliers in the level-set equation. Instead higher order polynomial approximations were used in the finite element method to increase accuracy, which increases the computational cost. In \cite{MaitreEtAl2012}, an other approach was used in the level-set context. In particular, a simple elastic force was introduced to penalize local stretching. This method requires a large elastic coefficient to generate nearly inextensible membranes that can introduce time step restrictions for stability. 

In \cite{Kimetal_JCP_2010,KimLai2012,HuEtAl2014}, a single-fluid model is also used with a Lagrange multiplier to enforce inextensibility but the scheme is implemented using a penalty immersed boundary method (iPB) in 2D and axisymmetric flows. In this approach, the interface is represented by two curves one of which moves with the fluid while the other moves elastically and under the influence of bending forces. The two curves are linked by stiff springs, which provide the only forces in the fluid. This approach enables the system for the fluid flow and the elastic and bending forces to be decoupled, which is in the same spirit as the method in \cite{Salacetal_JCP_2011,Salacetal_JFM_2012}. In principle, the method should converge to the original inextensible model as the spring stiffness tends to infinity, although this was not demonstrated and numerically large stiffnesses can introduce severe time step restrictions for stability.




Single-fluid models implemented using the phase field method were presented in \cite{Bibenetal_PRE_2003,Bibenetal_PRE_2005,MaitreEtAl2012}. In this approach, a Lagrange multiplier is introduced and is assumed to satisfy an advection-reaction equation where the advective time derivative is proportional to the surface divergence of the velocity field. The constant of proportionality is referred to as a tension-like parameter $T$. To ensure stability, additional diffusion is introduced which smooths out strong local variations in $\nabla_\Gamma\cdot{\bf v}$. As shown in the asymptotic analysis in \cite{Bibenetal_PRE_2005}, and further discussed in \cite{Jametetal_PRE_2007}, inextensibility in this approach is only fulfilled in the limit $T\to\infty$ where in practice, $T\sim\epsilon^{-1}$ and $\epsilon$ is proportional to the thickness of the diffuse interface, which is taken to zero. Thus, for finite $\epsilon$, the interface is not fully inextensible. The convergence of the method as $\epsilon\to 0$ was not demonstrated numerically.

Each of the methods discussed above has advantages and disadvantages. However, a common feature is that all the methods require various forms of regularization to implement the dynamics and to enforce the inextensibility of the vesicle membrane to some degree. As we demonstrate here, the dynamics of the vesicle can be very sensitive to the accuracy to which the inextensibility condition is modeled. Thus, there is still a need to develop models for which the accuracy of the inextensibility constraint can be explicitly controlled and for which convergence can be demonstrated. 

Accordingly, in this paper we present a new diffuse interface model for the dynamics of inextensible vesicles in a viscous fluid with inertia. A new feature of this work is the implementation of the local inextensibility condition in the diffuse interface context. As in the other methods described above, local inextensibility is enforced by using a local Lagrange multiplier, which provides the necessary tension force at the interface. However, to solve for the local Lagrange multiplier, we introduce a new equation whose solution essentially provides a harmonic extension of the local Lagrange multiplier off the interface while maintaining the local inextensibility constraint near the interface. To make the method more robust, we develop a local relaxation scheme that dynamically corrects local stretching/compression errors thereby preventing their accumulation. Asymptotic analysis is presented that shows that our new system converges to a relaxed version of the inextensible sharp interface model. This is also verified numerically. Although the model does not depend on dimension, we present numerical simulations only in 2D. To solve the 2D equations numerically, we develop an efficient algorithm combining an operator splitting approach with adaptive finite elements where the Navier-Stokes equations are implicitly coupled to the diffuse interface inextensibility equation. 

%


The outline of the paper is as follows. In Sec. \ref{phase field model}, the new diffuse interface models are derived. In Sec. \ref{sec:asymptotics}, a matched asymptotic analysis of the diffuse models is presented. In Sec. \ref{numerical methods}, the spatiotemporal discretization of the system is discussed. In Sec. \ref{numerical results}, numerical results are presented that demonstrate the convergence of the diffuse interface method as the interface thickness $\epsilon\to 0$ and that errors in enforcing local inextensibility may accumulate and lead to large differences in the dynamics in the tumbling regime and differences in the inclination angle of vesicles in the tank-treading regime. The local relaxation algorithm  is shown to effectively prevent this accumulation by driving the system back to its equilibrium state when errors in local inextensibility arise. In Sec. \ref{conclusions}, we present conclusions and discuss future work.

\section{Phase field/Diffuse interface models}
\label{phase field model}

The phase field method, also known as the diffuse interface method, introduces an auxiliary field $\phi$ that distinguishes the vesicle interior from the exterior. The vesicle boundary is modeled by a narrow, diffuse layer. An equation is posed for the phase field function $\phi$, which is nonlinearly coupled to the fluid equations. Near the interface, $\phi$ can be approximated by
\begin{equation}
 \phi(t,\mathbf{x}):=\tanh\left(\frac{-r(t,\mathbf{x})}{\sqrt{2}\epsilon}\right) \label{eq:phasefield}
\end{equation}
where $\epsilon$ characterizes the thickness of the diffuse interface and $r(t,\mathbf{x})$ denotes the signed-distance function between $\mathbf{x} \in \Omega$ and its nearest point on $\Gamma(t)$. Taking $r$ to be negative inside the vesicle, we label the inside with $\phi\approx 1$ and the outside with $\phi\approx -1$. The interface $\Gamma(t)$ is implicitly defined by the zero level set of $\phi$. 


Consider a diffuse interface version of the nondimensional Helfrich energy \cite{Duetal_Nonl_2005}
\begin{eqnarray}
\cal{E}(\phi) & = & \int_\Omega \frac{1}{2\epsilon}\frac{1}{\text{ReBe}} \left(\epsilon \Delta\phi-\frac{1}{\epsilon}(\phi^2-1)(\phi+H_0)\right)^2\; d\Omega. \label{diffuseEnergy}
\end{eqnarray}
where the Reynolds number is $\text{Re} = \rho_2 V L/\nu_2$, where $\rho_2$ and $\nu_2$  are the density and viscosity of the matrix fluid (the fluid outside the vesicle) and $L$ and $V$  are characteristic length and velocity scales. The bending capillary number is $\text{Be} = 4\sqrt{2}\nu_2 L^2 V/3b_N$, where $b_N$ is the bending stiffness. The scaling factor $4\sqrt{2}/3$ arises from the choice of the double-well potential $\left(\phi^2-1\right)\left(\phi+H_0\right)$ contained in Eq. \eqref{diffuseEnergy} and is chosen to match the sharp interface energy in the thin interface limit. For example, 
in \cite{Duetal_Nonl_2005} a formal convergence analysis as $\epsilon \to 0$ is performed to show that the diffuse interface energy in Eq. (\ref{diffuseEnergy}) tends to the nondimensional form of the sharp interface energy  in Eq. (\ref{eq:Helfrich_sharp}). This approach differs from the treatment in \cite{Bibenetal_PRE_2003} where the diffuse interface version of the Helfrich energy is the extension of the sharp interface energy in Eq. (\ref{eq:Helfrich_sharp}) off the interface into the whole domain $\Omega$ with the curvature and normal vector being calculated as $H = - \nabla \cdot \mathbf{n}$ and $\mathbf{n} = \nabla \phi / |\nabla \phi|$, respectively.

\subsection{Global surface area constraint: Model A}
 
A thermodynamically consistent phase field approach to model the dynamics of vesicles in a viscous fluid was proposed in \cite{Duetal_DCDS_2007,DuEtAl}. In this approach, spatially constant Lagrange multipliers were introduced to enforce volume and total (global) surface area conservation, and bending forces obtained variationally from the energy in Eq. (\ref{diffuseEnergy}) were included. The resulting nondimensional Navier-Stokes system is
\begin{eqnarray}
\rho (\partial_{t} \mathbf{v} + \mathbf{v} \cdot \nabla \mathbf{v}) + \nabla p - \frac{1}{\text{Re}} \nabla \cdot (\nu \mathbf{D} ) \!\!&=&\!\! g \nabla\phi -\lambda_{global} f \nabla\phi \nonumber \\
& &
+ \lambda_{volume}\nabla\phi,   \label{eq:navierStokesNonD1}\\
\nabla \cdot \mathbf{v} \!\!&=&\!\! 0, \label{eq:navierStokesNonD2}
\end{eqnarray}
where $\lambda_{global}$ and $\lambda_{volume}$ are the  Lagrange multipliers and the terms on the right hand side of Eq. (\ref{eq:navierStokesNonD1})  are the excess forces due to bending, global surface area conservation and volume conservation respectively. Further,
\begin{eqnarray}
g & = & \frac{1}{ \text{Re}\text{Be}} \left(\Delta f_c-\frac{1}{{\epsilon}^2}(3\phi^2+2H_0\phi-1) f_c\right), \label{g term}\\
f_c & = & \epsilon\Delta\phi-\frac{1}{{\epsilon}}(\phi^2-1)(\phi+H_0), \label{f_c term} \\
 f & =& \epsilon\Delta\phi-\frac{1}{{\epsilon}}(\phi^2-1)\phi. \label{f term}
 \end{eqnarray}
The evolution of $\phi$ is given by the dimensionless nonlinear advection-diffusion equation 
\begin{equation}
\partial_{t} \phi + \mathbf{v}\cdot \nabla \phi   = -\eta ( g - \lambda_{global} f + \lambda_{volume} ),\label{eq:phasefieldkonv}
\end{equation}
where $\eta>0$ is a small parameter. The density and viscosity are modeled as
 $\rho = \rho(\phi) = 0.5 (\phi + 1)\rho_1/\rho_2 + 0.5(1-\phi)$ and   $\nu = \nu(\phi) = 0.5 (\phi + 1)\nu_1/\nu_2 + 0.5 (1-\phi) $, respectively (see also  \cite{Bibenetal_PRE_2005,Salacetal_JCP_2011}). 
The Lagrange multipliers $\lambda_{volume}$ and $\lambda_{global}$ follow from the constraints
\begin{eqnarray*}
\frac{d}{dt} \mathcal{V}(\phi) = \frac{d}{dt} \int_\Omega \frac{1}{2} (\phi +1) \; d\Omega \!&=&\! 0 \qquad \mbox{(volume constraint)} \\
\frac{d}{dt} \mathcal{A}(\phi) = \frac{d}{dt} \int_\Omega \frac{\epsilon}{2} |\nabla \phi|^2 + \frac{1}{4 \epsilon} (\phi^2 - 1)^2) \; d\Omega \!&=&\! 0. \qquad \mbox{(global area constraint)} 
\end{eqnarray*}
Using the evolution equation for $\phi$,  the system to be solved for $\lambda_{volume}$ and $\lambda_{glo al}$ reads
\begin{eqnarray}
\lambda_{volume} \int_\Omega \; d \Omega + \lambda_{global} \int_\Omega f \; d \Omega \!&=&\!  \int_\Omega  g \; d\Omega,\label{eq:lagrange1} \\
\lambda_{volume} \int_\Omega f \; d \Omega + \lambda_{global} \int_\Omega f^2 \; d \Omega \!&=&\!  \int_\Omega  (\frac{1}{\eta} \mathbf{v} \cdot \nabla \phi + g) f \; d\Omega, \label{eq:lagrange2} 
\end{eqnarray}
which must be solved together with Eqs. (\ref{eq:navierStokesNonD1}), (\ref{eq:navierStokesNonD2}) and (\ref{eq:phasefieldkonv}). Because of the accumulation of errors, \cite{Duetal_JCP_2006} suggested that additional relaxation terms be added to the equations, which was found to improve accuracy. That is, the terms $\frac{1}{2 \tau} (\mathcal{V}_0-\mathcal{V}(\phi))$ and $\frac{1}{2 \tau} (\mathcal{A}_0-\mathcal{A}(\phi))$ are added to the right hand sides of Eqs. (\ref{eq:lagrange1}) and (\ref{eq:lagrange2}), respectively,  where $\mathcal{V}_0$ and $\mathcal{A}_0$ denote the desired volume and area. The relaxation parameter is the inverse of the time step size $\tau$.


\subsection{Local inextensibility constraint: Model B}

To enforce the local inextensibility constraint in the phase-field model, we propose a modification of the flow problem in Model A. In particular, we introduce spatially varying Lagrange multiplier $\lambda_{local}$, which introduces tension forces along the interface. These tension forces take the form $\nabla \cdot (\delta_\epsilon \mathbf{P} \lambda_{local})$, where $\mathbf{P} = \mathbf{I} - \mathbf{n} \otimes \mathbf{n}$, with $\mathbf{n} = -\nabla \phi / |\nabla \phi|$, is the tangential projection operator and $\delta_\epsilon = 0.5 |\nabla \phi|$ is a diffuse interface approximation of the surface delta function.

The nondimensional Navier-Stokes equation thereby becomes
\begin{eqnarray}
\!\!\!\!\!\!\!\rho (\partial_{t} \mathbf{v} + \mathbf{v} \cdot \nabla \mathbf{v}) +\nabla p - \frac{1}{\text{Re}} \nabla \cdot (\nu \mathbf{D} ) \!\!&=&\!\!  \nabla \cdot (\delta_\epsilon \mathbf{P} \lambda_{local})+g \nabla\phi   \nonumber \\
&& -\lambda_{global} f \nabla\phi + \lambda_{volume}\nabla\phi,   \label{eq:navierStokesNonD3}\\
\nabla \cdot \mathbf{v} \!\!&=&\!\! 0, \label{eq:navierStokesNonD4}
\end{eqnarray}
where we have also retained the volume and global surface area Lagrange multipliers, which we found to help improve the accuracy of the method (e.g., the inclusion of the volume and global surface area constraints means that $\lambda_{local}$ is decreased in magnitude compared to the case where the constraints are not included). 
The evolution equation for $\phi$ as well as the system to determine
$\lambda_{global}$ and $\lambda_{volume}$ remain as before.

The inextensibility constraint $\nabla_\Gamma \cdot \mathbf{v} = \mathbf{P} : \nabla \mathbf{v} = 0$ on $\Gamma$ is extended off $\Gamma$ into the whole domain $\Omega$, following the diffuse domain approach \cite{Raetzetal_CMS_2006,LiEtAlDDM,Teigenetal}. The idea is to perform an extension of the equation in order to solve for $\lambda_{local}$ in the whole domain, without extending the inextensibility constraint away from the interface. In particular, we take
\begin{equation}
\xi \epsilon^2 \nabla \cdot (\phi^2 \nabla \lambda_{local}) + \delta_\epsilon \mathbf{P} : \nabla \mathbf{v} = 0 \label{eq:navierStokesNonD5},
\end{equation}
where $\xi > 0$ is a parameter independent of $\epsilon$.  Eq. (\ref{eq:navierStokesNonD5}) reduces to $\Delta \lambda_{local} = 0$ away from $\Gamma$, since $\phi^2 \approx 1$ and $\delta_\epsilon \approx 0$, and becomes $\mathbf{P} : \nabla \mathbf{v} = 0$ near $\Gamma$, where $\delta_\epsilon$ is large and $\phi^2 \approx 0$. Thus, this effectively provides a harmonic extension of $\lambda_{local}$ off $\Gamma$ while maintaining the local inextensibility constraint near $\Gamma$. An asymptotic analysis is given in Sec.\ref{sec:asymptotics}, which shows convergence of Eq. ({\ref{eq:navierStokesNonD5}) as $\epsilon \rightarrow 0$ to the original sharp interface inextensibility constraint. We note that the tension force term in the Navier-Stokes equation expands to $\nabla \cdot (\delta_\epsilon \mathbf{P} \lambda_{local})=\delta_\epsilon (\nabla_\Gamma \lambda_{local} - \lambda_{local} H \mathbf{n})$, which is exactly the body force term used in \cite{Bibenetal_PRE_2003,Beaucourtetal_PRE_2004,Bibenetal_PRE_2005,Salacetal_JCP_2011,Laadharietal_PF_2012}.

Our approach differs from that taken in the phase field method used in  \cite{Bibenetal_PRE_2003,Beaucourtetal_PRE_2004,Bibenetal_PRE_2005,MaitreEtAl2012}, where the evolution equation 
$\partial_t \lambda_{local} + \mathbf{v} \cdot \nabla \lambda_{local} = \beta \Delta \lambda_{local} + T \mathbf{P} : \nabla \mathbf{v}$ was used instead of Eq. (\ref{eq:navierStokesNonD5}). In this equation, $T$ is interpreted as a tension-like constant that effectively controls the inextensibility of the membrane and the diffusion is only added for regularization purposes, with a small parameter $\beta > 0$. Note that if $T\to\infty$ (e.g., $T\sim 1/\epsilon$), then the inextensibility condition is enforced throughout the whole domain, which is unlike the formulation considered here. Further, unlike the case here, the additional Lagrange multipliers for volume and global area conservation were not considered. 


\subsection{Local inextensibility constraint with relaxation: Model C}

\noindent
As occurs with the global surface area constraint \cite{Duetal_JCP_2006}, solving Eq. ({\ref{eq:navierStokesNonD5}) may introduce small errors at each time step due to the regularization term (first term on the left hand side). Such errors may accumulate over time and may lead to spurious local stretching of the membrane. Hence, it would be desirable to have a local mechanism to correct these errors and drive a slightly stretched surface back to equilibrium. Such relaxation mechanisms were used in sharp interface models of flexible fibers evolving in a Stokes flow \cite{TornbergShelley2004}. Here, we present a local relaxation mechanism in the diffuse interface context. 

We introduce a variable $c$ to measure local stretching of the interface. Taking $c$ to evolve by the surface mass conservation equation:
\begin{eqnarray}
\partial_t c + {\bf v}\cdot\nabla c + c \nabla_\Gamma \cdot {\bf v} \!\!&=&\!\! 0 
 \qquad  \mbox{on $\Gamma$}, 
\label{evolution c sharp adv}
\end{eqnarray}
and setting the initial value $c(\bf{x},0) =1$, locations where $c$ deviates from $1$ represent regions of compression ($c>1$) and stretching ($c<1$). For numerical purposes we introduce additional diffusion along the interface
\begin{eqnarray}
\partial_t c + {\bf v}\cdot\nabla c + c \nabla_\Gamma \cdot {\bf v} \!\!&=&\!\! \theta \Delta_\Gamma c
 \qquad  \mbox{on $\Gamma$} 
\label{evolution c sharp}
\end{eqnarray}
with a small parameter $\theta > 0$. Restricting the diffusion to the interface ensures no interference with the bulk.

We use a version of Hooke's law to relax the local changes in interfacial area. In particular, we require that the strength of the relaxation is proportional to the amount of local stretching and compression. Accordingly, we take 
 $\nabla_\Gamma \cdot {\bf v} = \zeta (c-1)/c$, where $\zeta > 0$ is a constant controlling the strength of the relaxation. As we will see later in the diffuse interface model, a good choice for $\zeta$ is the inverse of the time step size. As long as $c = 1$ we have the original inextensibility condition $\nabla_\Gamma \cdot {\bf v} = 0$.
 

Within the domain formulation we replace Eq. \eqref{eq:navierStokesNonD5} by
\begin{equation}
\xi \epsilon^2 \nabla \cdot (\phi^2 \nabla \lambda_{local}) + \delta_\epsilon \mathbf{P} : \nabla \mathbf{v} = \zeta \frac{c-1}{c}\delta_\epsilon,
 \label{eq:navierStokesNonD5 relaxed}
\end{equation}
where the concentration $c$ satisfies a diffuse interface version of Eq. \eqref{evolution c sharp},  
\begin{eqnarray}
\partial_t c + {\bf v}\cdot\nabla c + c {\bf P}:\nabla {\bf v} \!\!&=&\!\!  \theta \nabla \cdot (\mathbf{P} \nabla c),
\label{evolution c diffuse} 
\end{eqnarray}
e.g., see \cite{Raetzetal_CMS_2006}.
The complete model including relaxation consists of solving the Navier-Stokes equation \eqref{eq:navierStokesNonD3}, \eqref{eq:navierStokesNonD4} and \eqref{eq:navierStokesNonD5 relaxed} for $\mathbf{v}$, $p$ and $\lambda_{local}$, the surface conservation equation \eqref{evolution c diffuse}  for $c$,  the phase field equation \eqref{eq:phasefieldkonv} for $\phi$, and Eqs. \eqref{eq:lagrange1}-\eqref{eq:lagrange2} for the Lagrange multipliers $\lambda_{global}$ and $\lambda_{volume}$.

At first glance, Model C appears to be  similar to the approach presented in \cite{Bibenetal_PRE_2003,Beaucourtetal_PRE_2004,Bibenetal_PRE_2005}. However here, the evolution equation here is for $c$, rather than $\lambda_{local}$ as in  \cite{Bibenetal_PRE_2003,Beaucourtetal_PRE_2004,Bibenetal_PRE_2005}. Further, since $c$ serves only to  correct errors in local inextensibility, we find that in practice our approach is relatively insensitive to the unknown relaxation rate $\zeta$ and tends to minimize errors introduced by the non-physical diffusion of $c$.

\section{Asymptotic analysis} 
\label{sec:asymptotics}

In this section, we use matched asymptotic expansions to show that Eq. (\ref{eq:navierStokesNonD5 relaxed}) converges as $\epsilon\to 0$ to the relaxed version of the sharp interface inextensibility condition
\begin{equation}
\mathbf{P} : \nabla \mathbf{v} = \nabla_\Gamma\cdot\mathbf{v}=\zeta \frac{c-1}{c}
\label{sharp interface BC}
\end{equation}
on the membrane surface $\Gamma(t)$, and Eq. (\ref{evolution c diffuse}) converges to the corresponding sharp interface Eq. (\ref{evolution c sharp}). In this approach, we expand the variables in powers of the interface thickness $\epsilon$ in regions close to (inner expansion) and far (outer expansion) from the interface. The two expansions are matched in an intermediate region where both expansions are presumed to be valid (e.g., see \cite{Caginalp88,Pego88} for a general description of the procedure). Previous work \cite{Duetal_DCDS_2007,DuEtAl} can be used to show that the Navier-Stokes system in Models B and C converge to the sharp interface incompressible Navier-Stokes equations with jump conditions given in Eq. (\ref{sharp local constraint}).

\paragraph{Outer expansion} Away from $\Gamma(t)$, which is defined as the zero level-set of $\phi$, we assume that all variables have a regular expansion in $\epsilon$. For example, the local Lagrange multiplier can be written as $\lambda_{local}=\lambda_{local}^{(0)}+\epsilon\lambda_{local}^{(1)}+\dots$, and likewise for the other variables. Further, away from $\Gamma$, we have $\phi=\pm 1$ to all orders and so $\grad\phi=\mathbf{0}$ and $\mathbf{P}=\mathbf{I}$ to all orders. Define the outer regions to be $\Omega^{+}$, the exterior of the vesicle, and $\Omega^{-}$ the interior of the vesicle. Accordingly, plugging the expansions into the equations and matching powers of $\epsilon$, Eq.  (\ref{eq:navierStokesNonD5 relaxed}) becomes
\begin{equation}
\Delta\lambda_{local}^{(i)}=0,~~{\rm for}~i=0,~1,\dots~~~{\rm in}~~\Omega^{\pm}.
\label{outer equation}
\end{equation}
The leading order contribution from Eq. (\ref{evolution c diffuse}) is:
\begin{equation}
\partial_t c^{(0)} + \mathbf{v}^{(0)}\cdot\grad c^{(0)} =  \theta\Delta c^{(0)}~~~\rm{in}~~\Omega^\pm.
\label{evolution c diffuse outer}
\end{equation}
%
%

\paragraph{Inner expansion} Near $\Gamma(t)$, we introduce a local coordinate system 
\begin{equation}
  \mathbf{x}(\mathbf{s},z;\epsilon) = \vct X(\mathbf{s};\epsilon)
    + \epsilon z\mathbf{n}(\mathbf{s};\epsilon),
\end{equation}
where $\mathbf{X}(\mathbf{s};\epsilon)$ is a parametrization of the interface, $\mathbf
{n}(\mathbf{s};\epsilon)$ is the interface normal vector that points out of the vesicle into $\Omega^+$,
$z$ is the stretched variable
\begin{equation}
  z = \frac{r(\mathbf{x)}}{\epsilon},
\end{equation}
and $r$ is the signed distance from the point $\mathbf{x}$ to $\Gamma(t)$, which is taken to be negative inside the vesicle. We then assume
that all variables can be written as functions of $z$ and $\mathbf{s}$ and that in these coordinates the variables have regular expansions in $\epsilon$.
That is, for the velocity field
\begin{equation}
  \hat {\vct v}(z,\vct s;\epsilon) \equiv \mathbf{v}(\vct x;\epsilon)
  = \mathbf{v} (\vct X(\vct s;\epsilon) + \epsilon z\vct n(\vct s;\epsilon);\epsilon),
\end{equation}
and the inner expansion is
\begin{equation}
  \hat {\vct v}(z,\vct s;\epsilon) = \hat{ \vct v}^{(0)}(z,\vct s)
    + \epsilon \hat {\vct v}^{(1)}(z,\vct s)
    + \epsilon^2\hat {\vct v}^{(2)}(z,\vct s) + \cdots.
  \label{eq:inner}
\end{equation}
The definitions and expansions of $\hat \phi$, $\hat \lambda_{local}$ and $\hat c$ are analogous. Note that $\hat\phi^{(0)}(z,\vct s)=\rm{tanh}\left(-z/\sqrt{2}\right)$, which can be justified using the analysis in \cite{Duetal_DCDS_2007,DuEtAl}.  

\paragraph{Matching conditions} The inner and outer expansions are matched in a region where both expansions are valid. To obtain the matching conditions, we assume that there is a region of overlap
where both the expansions are valid, e.g. where $\epsilon z = \bigo 1$. In particular, if we evaluate the outer expansion in the inner
coordinates, this must match the limits of the inner solutions away from the
interface. This procedure provides boundary conditions for the outer equations. Summarizing the results for the velocity field (the matching conditions for the other fields are analogous) we have \cite{Pego88}
\begin{equation}
  \label{eq:match1}
  \zlimpm \hat{\vct v}^{(0)}(z,\vct s) = \mathbf{v}^{(0)}(\vct s),
\end{equation}
at leading order. At  the next order, we obtain
\begin{align}
  \label{eq:match2}
  \hat {\vct v}^{(1)}(z,\vct s) &= \mathbf{v}^{(1)}(\vct s)
    + z\ndot\grad \mathbf{v}^{(0)}(\vct s) + \smallo 1, 
\end{align}
as $z\to\pm\infty$, and so on.  The quantities on the right hand sides in Eqs. (\ref{eq:match1}) and (\ref{eq:match2}) are the limits from the interior ($\Omega^-$) and exterior ($\Omega^+$)
of  the vesicle. Here $\smallo 1$ means that the expressions approach equality when
$z\to\pm\infty$.  That is, $\smallo 1$ is defined such that if some function
$f(z) = \smallo 1$, then we have $\zlimpm f(z) = 0$. 

\paragraph{Analysis near $\Gamma$} In the local coordinate system, the
derivatives become
\begin{align}
\label{time deriv}
\partial_t&=-\frac{V}{\epsilon}\partial_z+\partial_t,\\
  \label{eq:grad-i}
  \grad &= \frac{1}{\epsilon}\vct n\partial_z + \gradG, \\
  \label{eq:lapl-i}
  \lapl &= \frac{1}{\epsilon^2}\partial_{zz}
           + \frac{H}{\epsilon}\partial_z + \laplG,
\end{align}
where $V$ is the normal velocity of $\Gamma$. Note that in Eq. (\ref{time deriv}) we have abused notation; what we mean here is  $\displaystyle{\partial_t c=-\frac{V}{\epsilon}\partial_z\hat{c}+\partial_t\hat{c}}$ and analogously for the other variables (e.g., see \cite{Caginalp88,Pego88}).

Define $\mathcal{P}=\mathbf{P}:\grad v$. It can be shown that the inner expansion of this term takes
the form
\begin{equation}
 \hat{\mathcal{P}}=\hat{\mathcal{P}}^{(0)}+\epsilon \hat{\mathcal{P}}^{(1)}+\dots,
 \label{inner expansion of stretching}
 \end{equation}
 where the leading term is given by
 \begin{equation}
\hat{\mathcal{P}}^{(0)}=\grad_\Gamma\cdot\hat{\vct v}^{(0)}.
\label{leading term of stretching}
\end{equation}
Eqs. (\ref{inner expansion of stretching}}) and (\ref{leading term of stretching}}) are justified in the Appendix. It is also shown in the Appendix that the leading order velocity field
\begin{equation}
\partial_z\hat{\vct v}^{(0)}=\mathbf{0}.
\label{leading order velocity field}
\end{equation}
Using this, together with the matching condition (\ref{eq:match1}) we conclude that the outer velocity $\mathbf{v}^{(0)}$ is continuous across the interface. Further, a straightforward calculation shows that
\begin{equation}
\hat\delta_\epsilon=-\frac{1}{2\epsilon}\hat\phi^{(0)}_z+\hat\delta_\epsilon^{(0)}+\epsilon \hat\delta_\epsilon^{(1)},
\label{expansion of delta fcn}
\end{equation}
where we do not present the specific forms of the higher order terms. 

At leading order $O(1/\epsilon)$, Eq. (\ref{eq:navierStokesNonD5 relaxed}) becomes:
\begin{equation}
\phi_z\hat{\mathcal{P}}^{(0)}=\zeta\frac{\hat{c}^{(0)}-1}{\hat{c}^{(0)}}\phi_z.
\label{leading order inner}
\end{equation}
Since $\phi_z<0$, we conclude that $\displaystyle{\hat{\mathcal{P}}^{(0)}=\nabla_\Gamma\cdot\hat{\vct v}^{(0)}=\zeta\frac{\hat{c}^{(0)}-1}{\hat{c}^{(0)}}}$. Taking the limit as $z\to\pm\infty$, using the matching condition and the continuity of the velocity, we obtain the inextensibility condition
\begin{equation}
\nabla_\Gamma\cdot\mathbf{v}^{(0)}=\zeta\frac{c^{(0)}-1}{c^{(0)}}~~~\rm{on}~~\Gamma(t),
\label{inextensibility}
\end{equation}
as claimed.

To analyze Eq. (\ref{evolution c diffuse}) in the inner variables, we use the fact that $\hat{\vct v}_z=0$ and that the interface moves with the fluid velocity at leading order: $V=\hat{\vct v}^{(0)}$, e.g., see \cite{Duetal_DCDS_2007,DuEtAl}. Then, Eq. (\ref{evolution c diffuse}) becomes
\begin{equation}
\partial_t\hat{c}^{(0)}+\hat{\vct v}^{(0)}\cdot\nabla_\Gamma\hat{c}^{(0)}+\hat{c}^{(0)}\nabla_\Gamma\cdot\hat{\vct v}^{(0)}=\theta\Delta_\Gamma\hat{c}^{(0)}.
\label{inner concentration eqn}
\end{equation}
Taking the limit $z\to\pm\infty$ and using the leading order matching condition (\ref{eq:match1}) we obtain
\begin{equation}
\partial_t{c}^{(0)}+\mathbf{v}^{(0)}\cdot\nabla_\Gamma c^{(0)}+{c}^{(0)}\nabla_\Gamma\cdot\mathbf{v}^{(0)}=\theta\Delta_\Gamma {c}^{(0)},
\label{outer limit of inner concentration eqn}
\end{equation}
the solution of which provides the boundary condition for Eq. (\ref{evolution c diffuse outer}).
Thus, putting everything together, we find that the leading order system is precisely the relaxed version of the 
locally inextensible, sharp interface Navier-Stokes model described earlier in Sec. \ref{introduction}. Of course, by setting $\zeta=0$, we obtain the locally inextensible sharp interface Navier-Stokes model.

\section{Numerical methods}
\label{numerical methods}

To  solve the system of equations numerically we split the time interval $I = [0,T]$ into equidistant time instants $0 = t_0 < t_1 < \ldots$ and define the time steps $\tau := t_{n+1}- t_n$. Of course, adaptive time steps may also be used. We define the discrete time derivative $d_t \cdot ^{n+1} := (\cdot ^{n+1} - \cdot ^{n})/ \tau$, where the upper index denotes the time step number. 
For Model C we choose the relaxation speed $\zeta=1/\tau$. One easily verifies that with this choice, errors in the inextensibility from the previous time step are approximately eliminated in the next time step.

The numerical approach for each subproblem is adapted from existing algorithms for the Navier-Stokes equations and the Helfrich model.  We solve the overall system using an operator splitting approach, with the Navier-Stokes equations being implicitly coupled  to the inextensibility constraint. The phase field variable is solved separately, as are the global Lagrange multipliers and the relaxation variable $c$.

We present here the time discretization of the inextensibility model with relaxation (Model C). At each time step we solve
\begin{enumerate}
\item The flow problem for $\mathbf{v}^{n+1}$, $p^{n+1}$ and $\lambda_{local}^{n+1}$:
\begin{align}
&\rho^n(d_t \mathbf{v}^{n+1} +\mathbf{v}^n\cdot\nabla \mathbf{v}^{n+1}) + \nabla p^{n+1}  \!\!- \frac{1}{\text{Re}} \nabla \cdot (\nu^n \mathbf{D}^{n+1}) - \nabla \cdot (\delta_\epsilon^n \mathbf{P}^n \lambda_{local}^{n+1}) \nonumber \\ 
&\qquad= g^n \nabla\phi^n -\lambda^n_{global} f^n \nabla\phi^n + \lambda^n_{volume}\nabla\phi^n, \\
&\nabla \cdot \mathbf{v}^{n+1} =0,\\
&\xi \epsilon^2 \nabla \cdot ((\phi^n)^2 \nabla \lambda_{local}^{n+1}) + \delta_\epsilon^n \mathbf{P}^n : \nabla \mathbf{v}^{n+1} = \tau^{-1} \frac{c^{n}-1}{c^n}\delta_\epsilon^n, \label{inextensibility equation discrete}
\end{align}
where 
$\rho^n = \rho(\phi^n)$, $\nu^n = \nu(\phi^n)$, $\mathbf{P}^n = \mathbf{I} - \frac{\nabla \phi^n \otimes \nabla \phi^n}{|\nabla \phi^n|^2}$ and $\delta_\epsilon^n = 0.5 |\nabla\phi^n|$.

\item The evolution equations for $\phi^{n+1}$, $g^{n+1}$, $f_c^{n+1}$ and $f^{n+1}$:
\begin{align}
&d_t\phi^{n+1} + \mathbf{v}^{n+1}\cdot \nabla \phi^{n+1}  = -\eta \left(g^{n+1} -\lambda_{global}^n f^{n+1} +\lambda_{volume}^n \right),\\
&g^{n+1}  = \frac{1}{\text{Re}\text{Be}}\left(\Delta f_c^{n+1}-\frac{1}{{\epsilon}^2}(3(\phi^{n+1})^2+2H_0\phi^{n+1}-1) f^{n+1}_c\right), \\
&f_c^{n+1}  = \epsilon\Delta\phi^{n+1}-\frac{1}{{\epsilon}}((\phi^{n+1})^2-1)(\phi^{n+1}+H_0),\\
&f^{n+1}  = \epsilon\Delta\phi^{n+1}-\frac{1}{{\epsilon}}((\phi^{n+1})^2-1)\phi^{n+1}.
\end{align}
We further linearize the nonlinear terms using a Taylor series expansion of order one, e.g. 
$((\phi^{n+1})^2-1)\phi^{n+1}=((\phi^n)^2-1)\phi^n+(3{(\phi^n)}^2-1)(\phi^{n+1}-\phi^n)$.
\item The equations for the Lagrange multipliers $\lambda^{n+1}_{volume}$ and $\lambda^{n+1}_{global}$:
\begin{align*}
&\lambda_{volume}^{n+1} \int_\Omega \; d \Omega + \lambda_{global}^{n+1} \int_\Omega f^{n+1} \; d \Omega =  \int_\Omega  g^{n+1}\; d\Omega + \frac{1}{2 \tau} (\mathcal{V}_0-\mathcal{V}(\phi^{n+1})), \\
&\lambda_{volume}^{n+1} \int_\Omega f^{n+1} \; d \Omega + \lambda_{global}^{n+1} \int_\Omega (f^{n+1})^2 \; d \Omega \\
&\qquad=  \int_\Omega  (\frac{1}{\eta} \mathbf{v}^{n+1} \cdot \nabla \phi^{n+1} +  g^{n+1}) f^{n+1}\; d\Omega +\frac{1}{2 \tau} (\mathcal{A}_0-\mathcal{A}(\phi^{n+1})).
\end{align*}
This system is solvable since the determinant of coefficients on the left hand side is positive, as long as $f^{n+1}$ is not a constant function. 

\item The advection-diffusion equation for the stretching variable $c^{n+1}$:
\begin{align}
\partial_t c^{n+1} + {\bf v}^{n+1}\cdot\nabla c^{n+1}+ c^{n+1} {\bf P}^{n+1}:\nabla {\bf v}^{n+1} &= \theta \nabla\cdot({\bf P}^{n+1}\nabla c^{n+1}).
\label{evolution c discrete} 
\end{align}

\end{enumerate}

\noindent
To solve the system without relaxation (model B), we omit the right hand side in Eq. \ref{inextensibility equation discrete}. For the global
area constraint (model A), we additionally omit the last equation of the flow problem in step 1 and set $\lambda_{local}^{n+1}=0$.
 
We use the adaptive finite element toolbox AMDiS \cite{amdis} for discretization in space, with the $P^2$/$P^1$ Taylor-Hood element for the flow problem, extended by a $P^2$ element for $\lambda_{local}$. For $\phi$ and $c$, $P^2$ elements are used. 
The resulting linear systems of equations are solved with UMFPACK \cite{umfpack}. The adaptive mesh refinement and coarsening are controlled by the phase field variable, for which a specified spatial resolution
at the interface that depends on $\epsilon$ is required. The choices for the numerical parameters $\eta$, $\xi$, $\zeta$ and $\theta$ are described in the next section.

\section{Numerical results}
\label{numerical results}

We conduct numerical tests to validate and to compare the results from models A, B and C. 
We focus on a single vesicle in shear flow in 2D, but the methods can also be used to simulate the interaction of many vesicles and can be extended to 3D. We prescribe ${\bf v}=(10,0)$ at the upper boundary and ${\bf v}=(-10,0)$ at the lower boundary of the domain $\Omega=[0,4]^2$. 
An open boundary conditions is used for ${\bf v}$ on the left and right boundary. 
We use homogeneous Neumann boundary conditions for $\lambda_{local}, f_c$  and  $\phi$, and the Dirichlet boundary condition $c=1$.
The initial vesicle is an ellipse, oriented in the $y$-direction, with major axis of length 2.5 and minor axis of length 1.0, placed in the center of the domain.
The initial velocity is zero. We also take $Be=20$, $H_0=0$, $\eta=0.1$, $\rho_1/\rho_2=1$, $\nu_1/\nu_2=10$, hence the viscosity is larger inside of the vesicle. 
Two different Reynolds numbers are used: $Re=1$ and $Re=1/200$ which correspond to the TT and TB regime, respectively. 
The regularization constant and relaxation rate are taken to be $\xi=1$, $\zeta=1/\tau$, respectively. The surface diffusion coefficient in the equation for $c$ is set to $\theta=0.01$. 
Unless otherwise stated, we use $\epsilon=0.03$ together with the minimum grid size $h=2^{-5}$ and the time step size $\tau=5.0e-4$.

There are two ways of measuring the inextensibility. One can either measure $\nabla_\Gamma\cdot {\bf v}$ at the interface or one can use the concentration variable $c$. In the latter case, the value of $(c-1)/c$ at the interface represents the local stretching accumulated over time while the former case measures the instantaneous stretching. Hence, to test the accuracy of our method we introduce the two test quantities:
\begin{align*}
E_{\bf v} &= \int_\Omega \epsilon^{-1}(1-\phi^2)^2 |\nabla_\Gamma\cdot{\bf v}|~d\Omega &\mbox{(instantaneous stretching),} \\ 
E_{c} &= \int_\Omega \epsilon^{-1}(1-\phi^2)^2 |(c-1)/c|~d\Omega & \mbox{(accumulated stretching).}
\end{align*}
Note that $\epsilon^{-1}(1-\phi^2)^2$ is a (scaled) diffuse interface approximation of the surface delta function.

\subsection{Convergence tests} 
\begin{figure}
\centering
\includegraphics[width=0.45\textwidth]{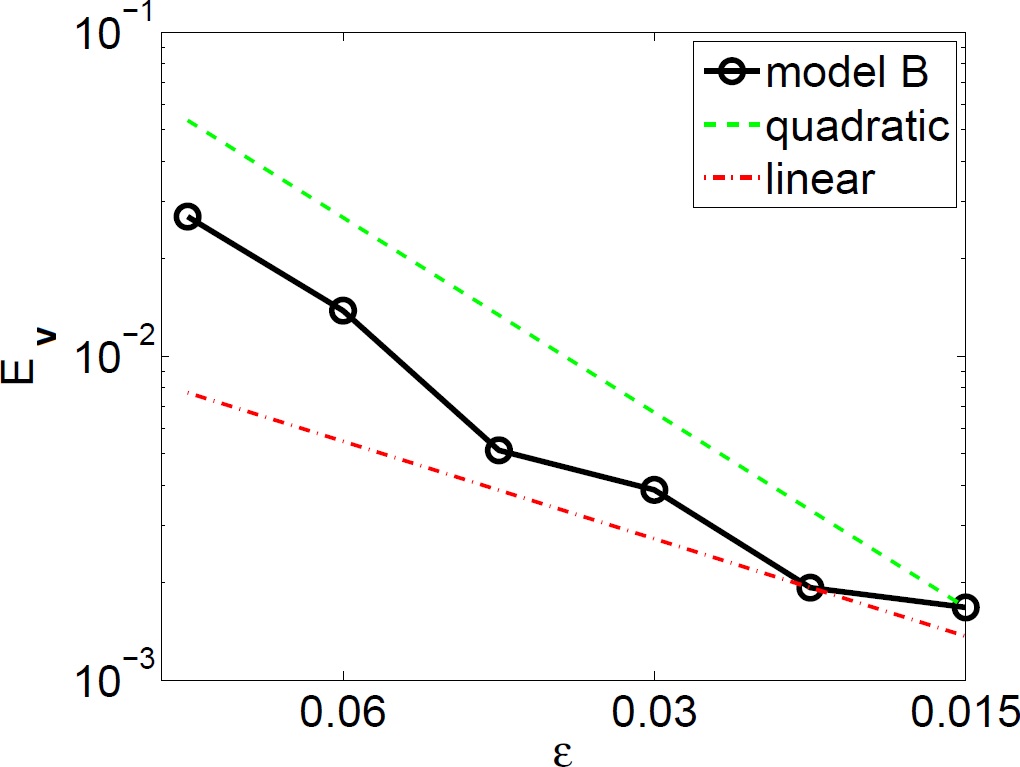}
\caption{
Convergence study showing super-linear decrease of the instantaneous stretching $E_{\bf v}$ as a function of the interface thickness $\epsilon$ for model B. (color online)
}
\label{fig:conv surf div}
\end{figure}

To validate the analytical results from Sec. \ref{sec:asymptotics} we conduct numerical convergence tests.
We use $Re=1$ and vary $\epsilon=0.848,0.06, 0.0424,0.03, 0.0212, 0.015$, where the grid is refined accordingly to have the same number of grid points across the interface. 
In Fig. \ref{fig:conv surf div} we display the error in the inextensibility condition measured using the instantaneous stretching $E_{\bf v}$ at the interface. 
The errors are calculated at the early time $t=0.025$ since a fair comparison between the results with different $\epsilon$ values is only possible if the vesicles are at similar positions. 
Results are only shown for Model B, and we find convergence rates between first and second order in $\epsilon$. As indicated by our asymptotic analysis in Sec. \ref{sec:asymptotics},
$E_{\bf v}$ is not expected to converge to 0 as $\epsilon\to 0$ for Model C, and in fact it does not (results not shown).

\begin{figure}
\centering
\includegraphics[width=0.45\textwidth]{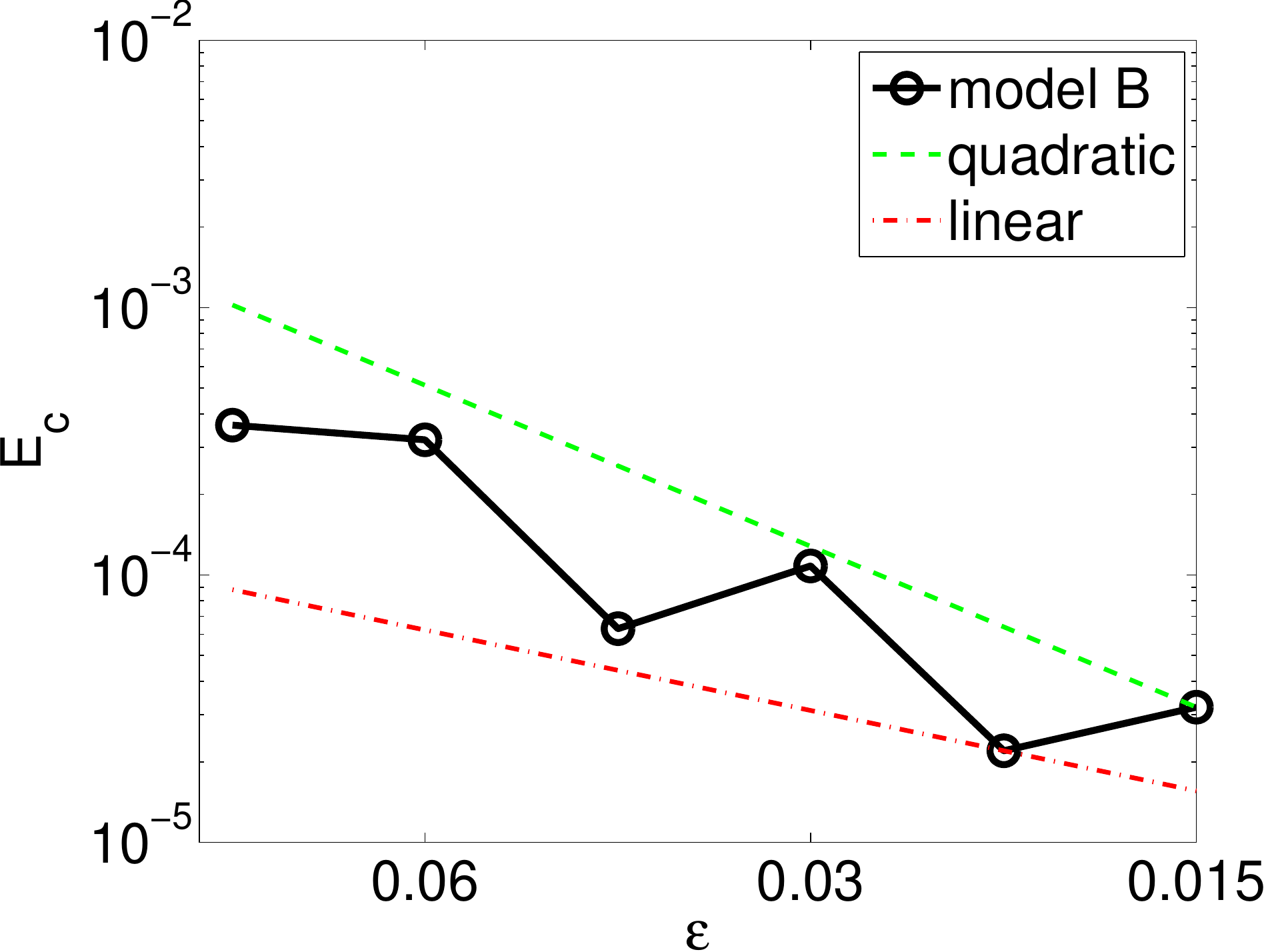}
\includegraphics[width=0.45\textwidth]{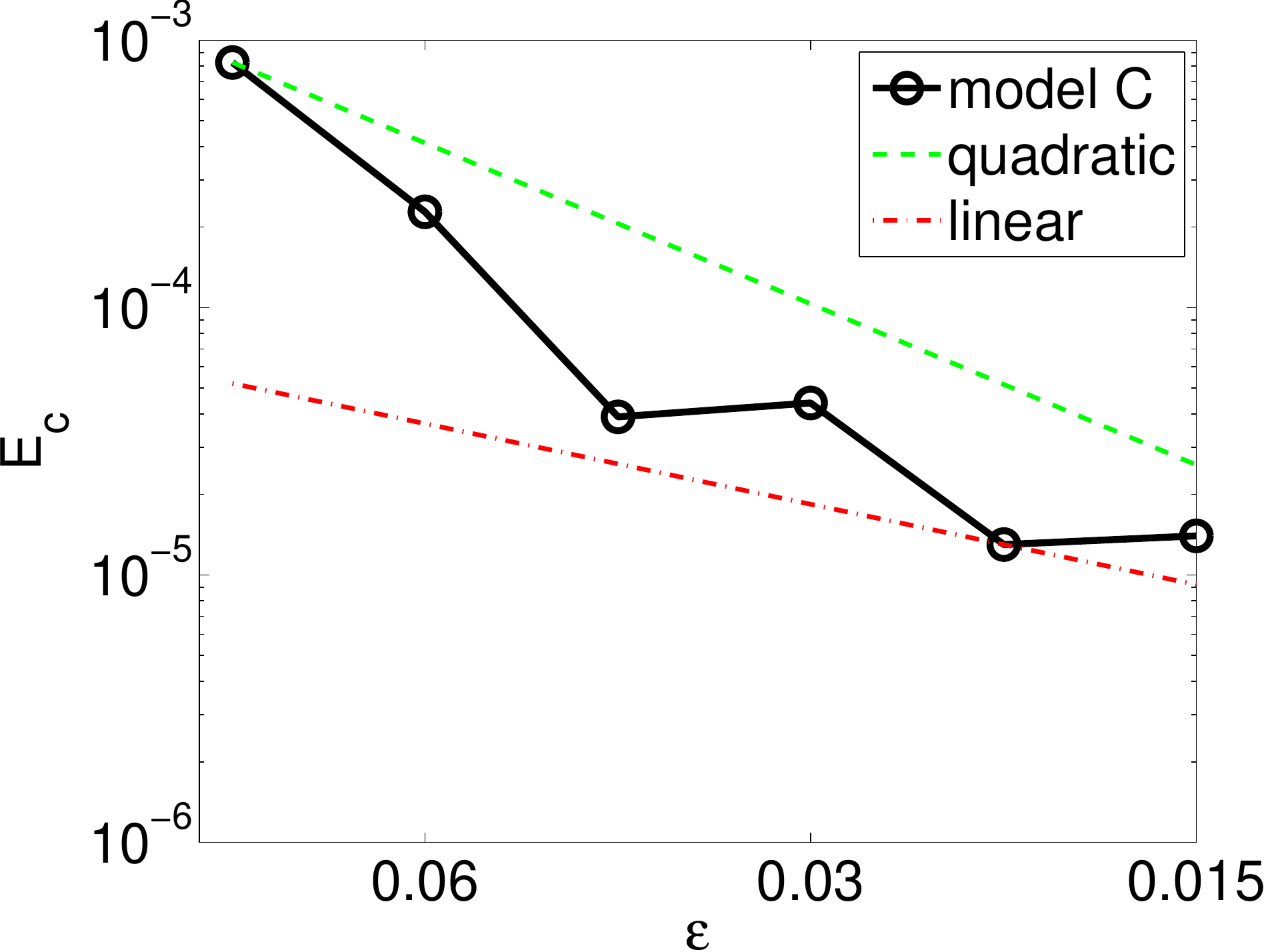}
\caption{Convergence study showing super-linear decrease of $E_c$ for decreasing $\epsilon$, for models B (left) and C (right). (color online)
}
\label{fig:conv stretching}
\end{figure}

In Fig. \ref{fig:conv stretching} the accumulated stretching $E_c$ is shown at $t=0.5$ for both Models B and C. Here, we also find convergence rates between first and second order in both models. 
Because the simulation time is short, the accumulated stretching is similar in both models, although the accumulated stretching is somewhat smaller in Model C as $\epsilon$ is decreased. In the next section, we show that simulating for longer times reveals a substantial difference in the accumulated stretching between the two models.


\subsection{Model comparison}
\begin{figure}
\centering
\includegraphics[width=0.45\textwidth]{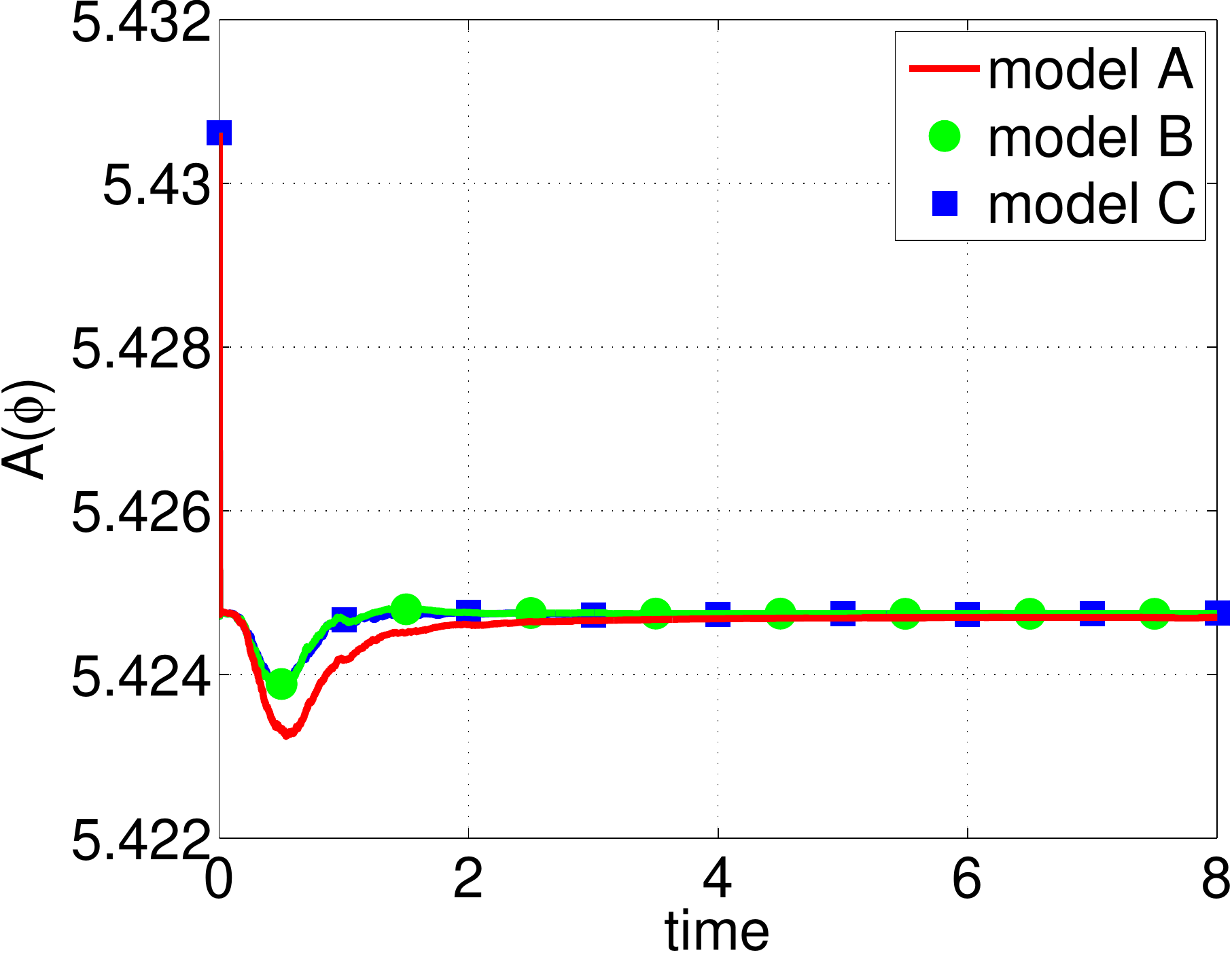}
\includegraphics[width=0.45\textwidth]{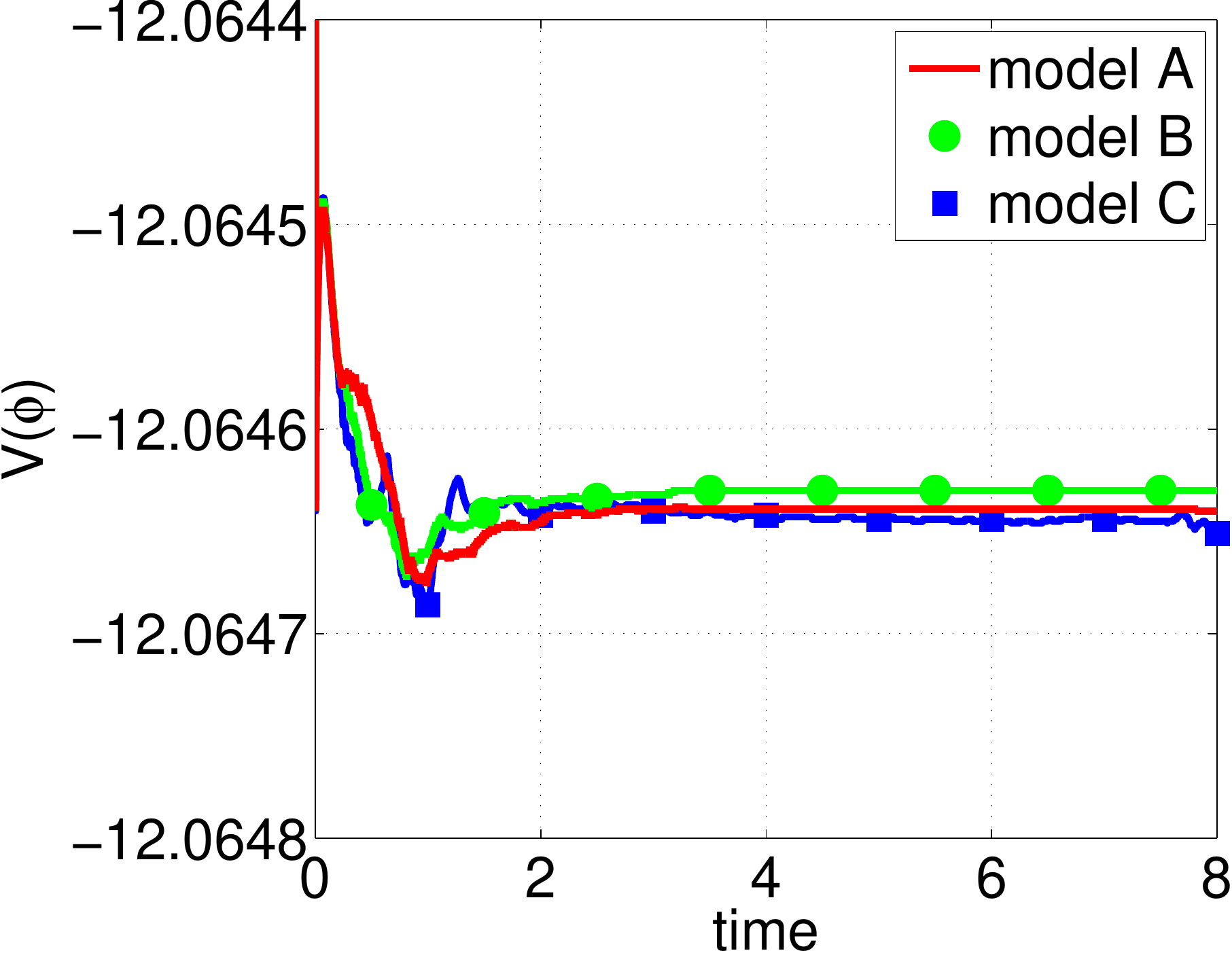}
\caption{Comparison for interface area (left) and vesicle volume (right) for $Re=1$. (color online)
}
\label{fig:model comparison volume and area}
\end{figure}

Next, we compare the results obtained from the different models. 
First, we validate the conservation of vesicle volume $V(\phi)$ and total interface area $A(\phi)$ for $Re=1$ in Fig. \ref{fig:model comparison volume and area}. 
The interface area is slightly better conserved by the use of the local inextensibility constraints in Models B and C. The slight drop in interfacial area around $t=0$ is due to the fact that the initial interface is not quite equilibrated (e.g., the initial interface profile is not represented by a hyperbolic tangent in the normal direction across the interface); equilibration occurs over the first few time steps. 
The vesicle volume is very well conserved by all the models, with the small variations at early times also being due to the equilibration of the interface.

\begin{figure}
\begin{tabular}{cccc}
  \includegraphics[width=0.21\textwidth]{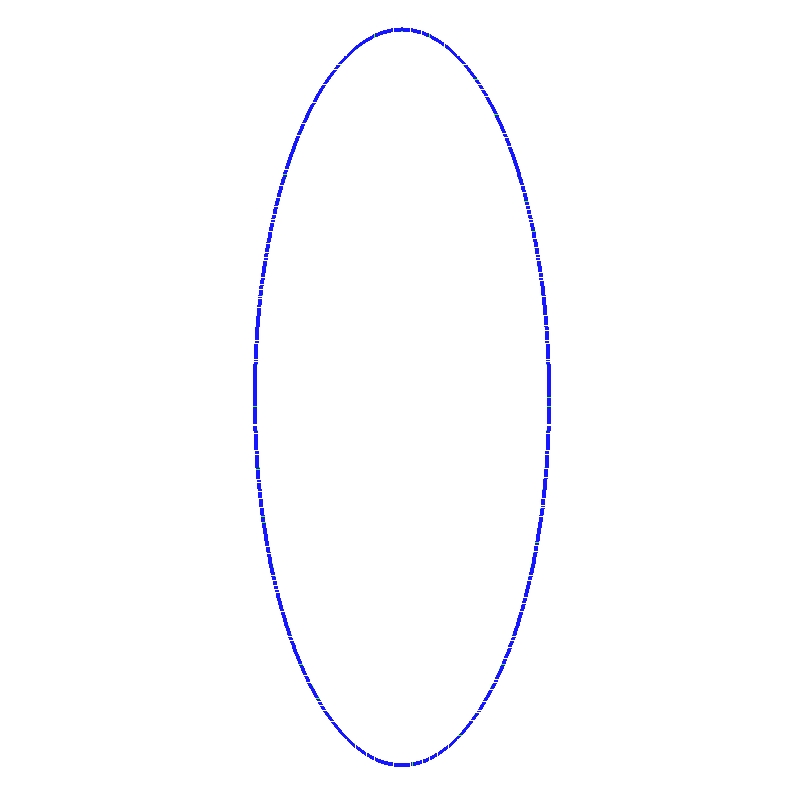} 
&  \includegraphics[width=0.21\textwidth]{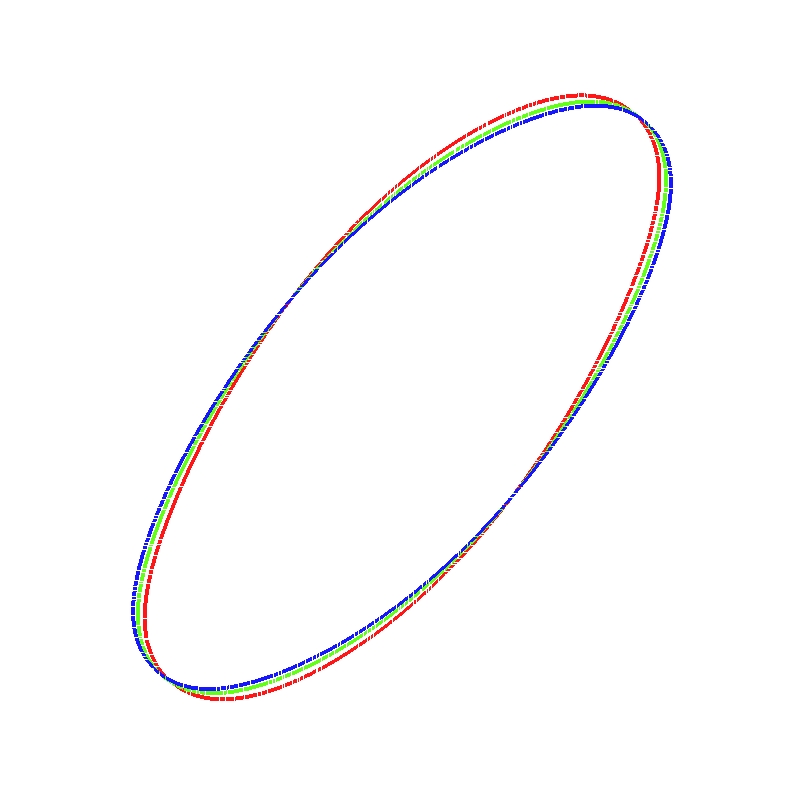}
&  \includegraphics[width=0.21\textwidth]{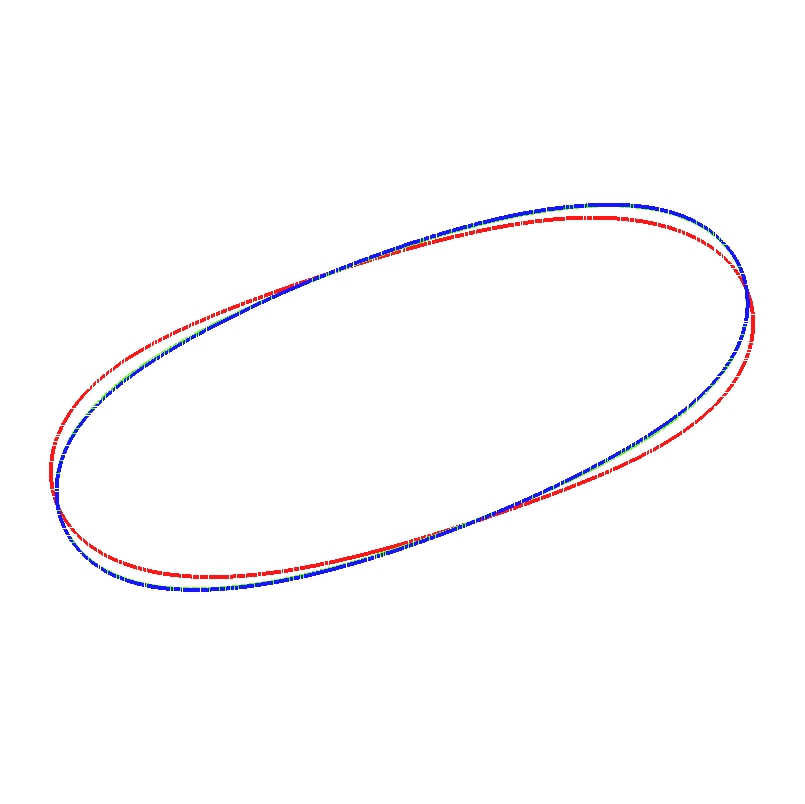}
&  \includegraphics[width=0.21\textwidth]{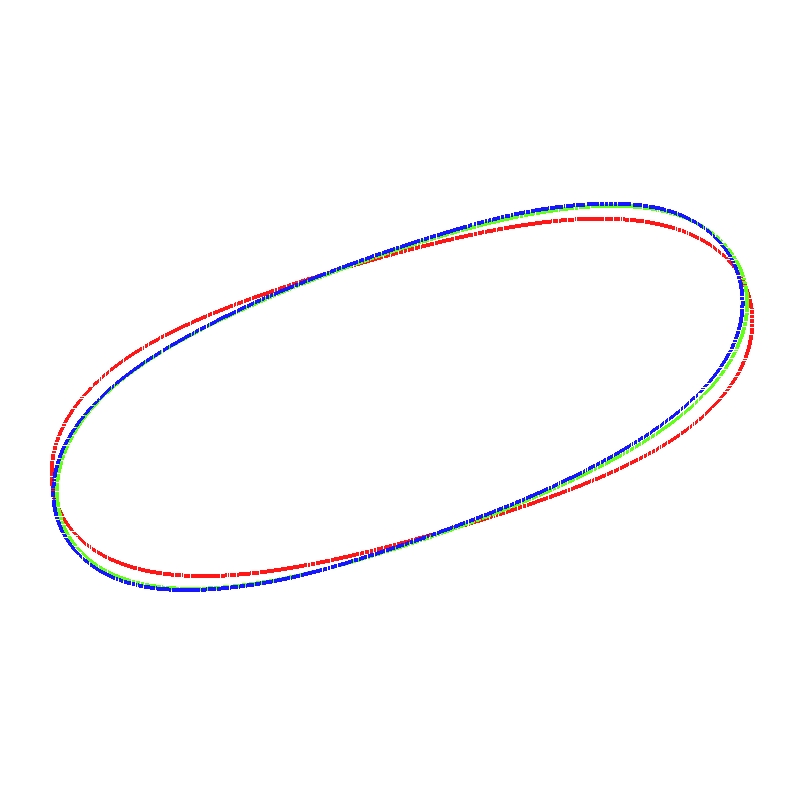}  \\
  $t=0$ &   $t=0.5$ &   $t=2.0$ &   $t=8.0$  \\\\      
  \includegraphics[width=0.21\textwidth]{images/snapshots/Re1t0.jpg}
&  \includegraphics[width=0.21\textwidth]{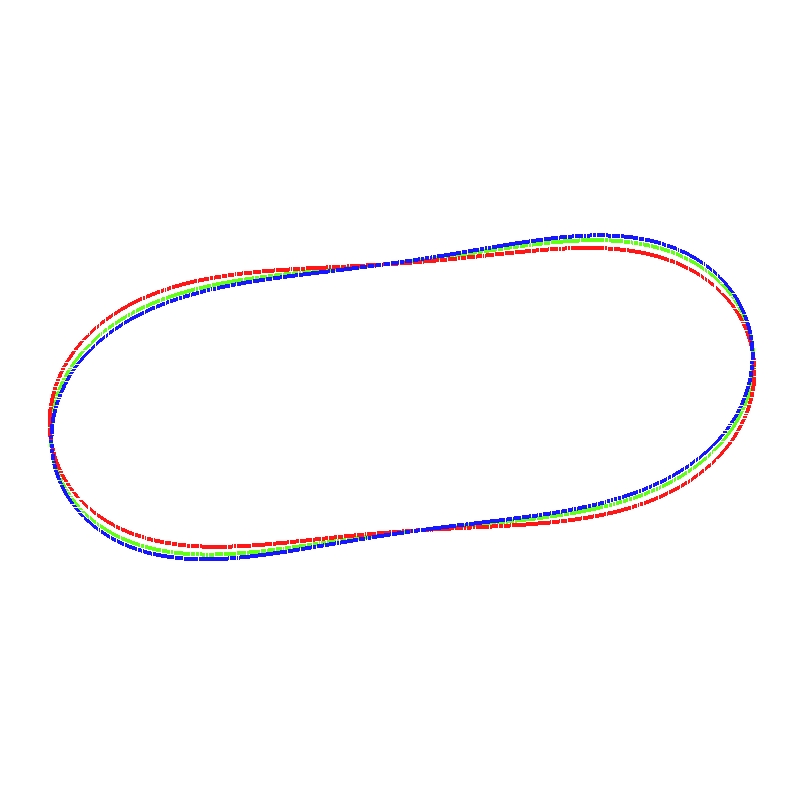}        
&  \includegraphics[width=0.21\textwidth]{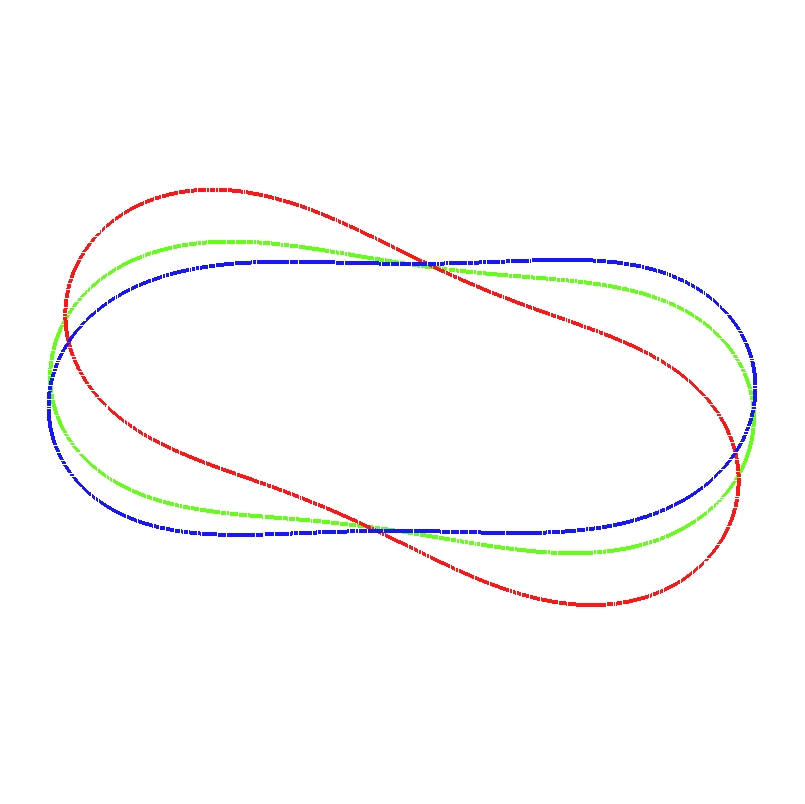}        
& \includegraphics[width=0.21\textwidth]{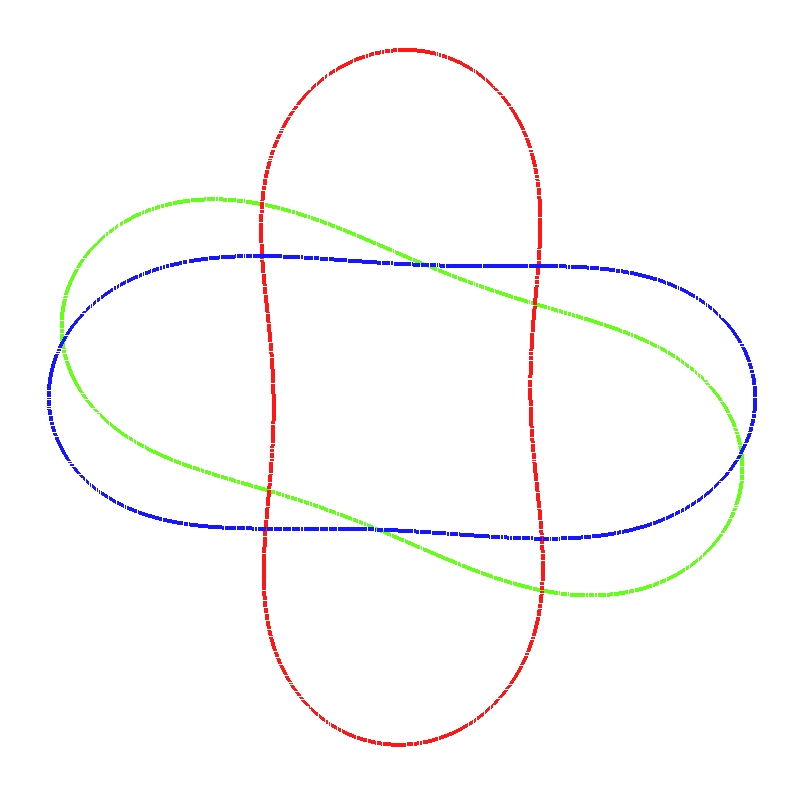}         \\
  $t=0$ &   $t=1.0$ &   $t=2.0$ &   $t=2.5$  \\\\
\end{tabular}      
  \caption{Time evolution of vesicles with $Re=1$ (top) and $Re=1/200$ (bottom) using Model A (red),
  Model B (green) and Model C (blue). The local inextensibility constraints in Models B and C tend to slow the rotation of the vesicle, which is particularly evident when $Re=1/200$.
 (Color online). }
  \label{fig: time evolutions}
\end{figure}

The time evolutions of vesicles obtained using the different models with
 $Re=1$ and $Re=1/200$ are shown in Fig. \ref{fig: time evolutions}-- the red corresponds to Model A, the green to Model B and the blue to Model C.  
When $Re=1$ (top graphs) the vesicle is in the TT regime and assumes a stationary state at around $t=2.0$ for all models. 
At small times (e.g. $t=0.5$) the local inextensibility constraints in Models B and C lead to a faster rotation and thus a smaller inclination angle of the vesicle . This effect is reversed for later times (e.g. $t=2.0, t=8.0$) and Model A leads to an approximately $4^\circ$ smaller inclination angle at the stationary state (see Fig. \ref{fig:model comparison angle} below). 


When $Re=1/200$ (bottom graphs) we find even larger differences between the models. 
The local inextensibility constraints in Models B and C delay the tumbling point significantly as well as the tumbling period. 
This is also seen in Fig. \ref{fig:model comparison angle} where the inclination angles of the vesicles are shown for the different models with $Re=1$ and $Re=1/200$. 

\begin{figure}
\centering
\includegraphics[width=0.45\textwidth]{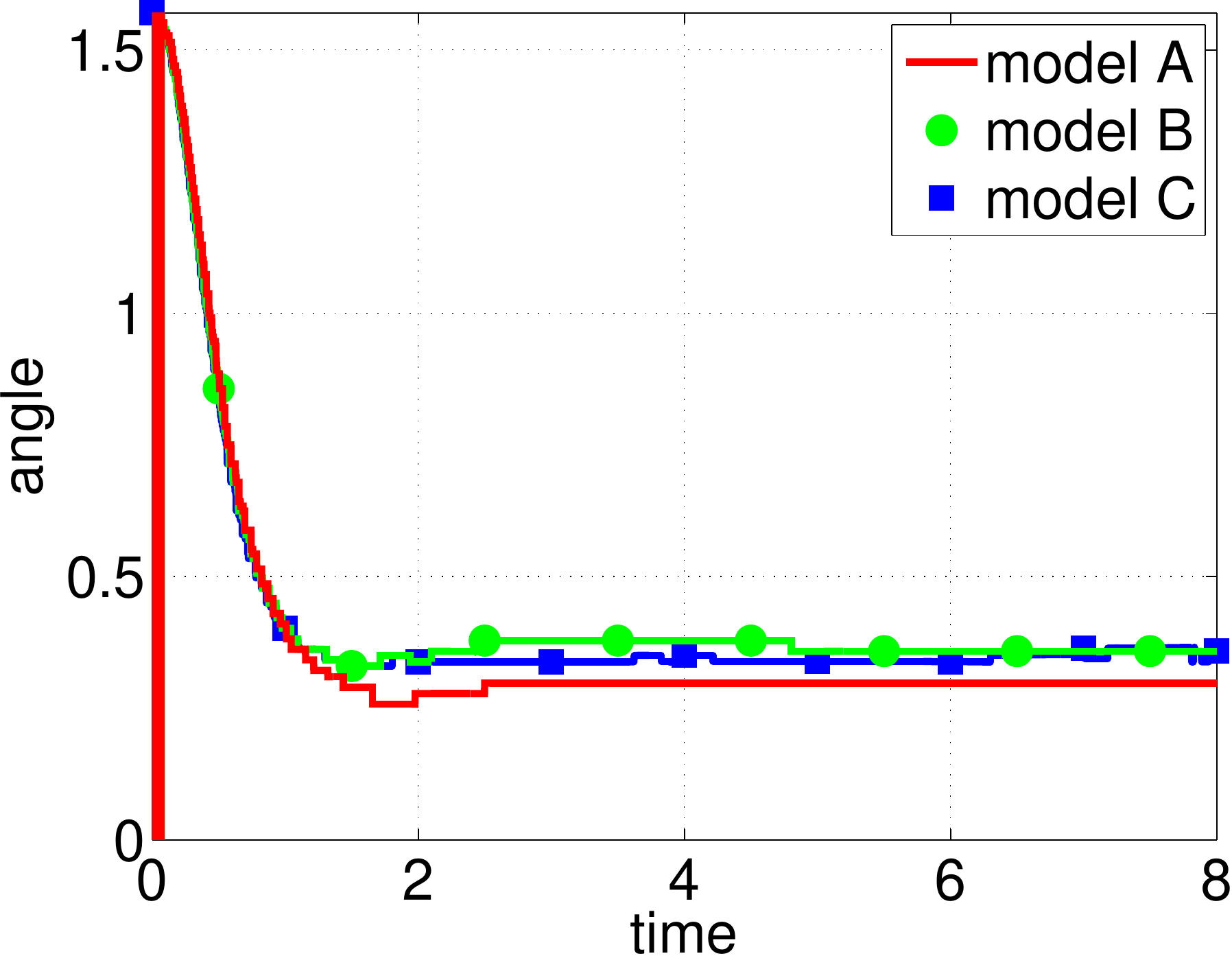}
\includegraphics[width=0.45\textwidth]{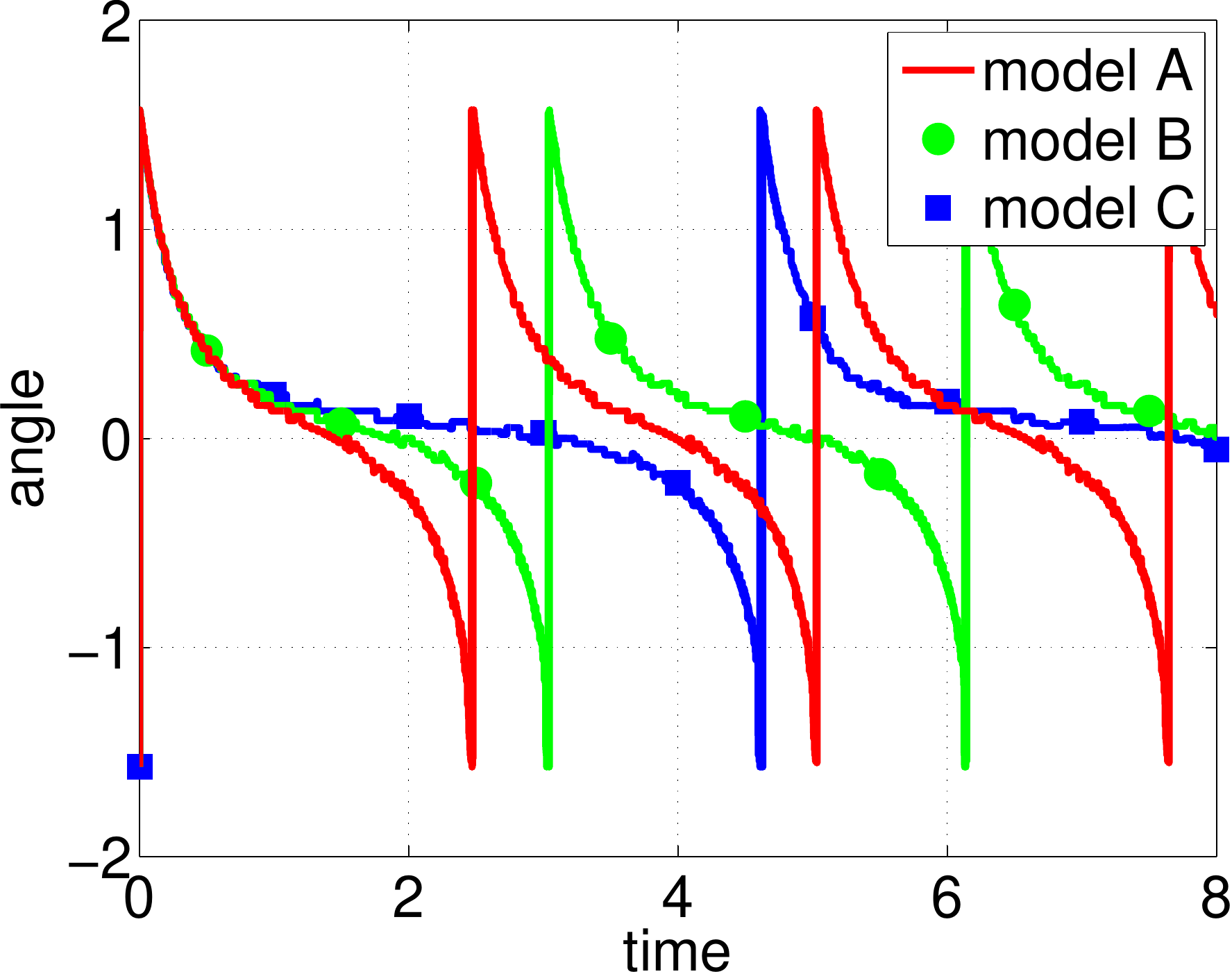}
\caption{Model comparison of the inclination angle for $Re=1$ (left) and $Re=1/200$ (right).
}
\label{fig:model comparison angle}
\end{figure}

\begin{figure}
\centering
\includegraphics[width=0.45\textwidth]{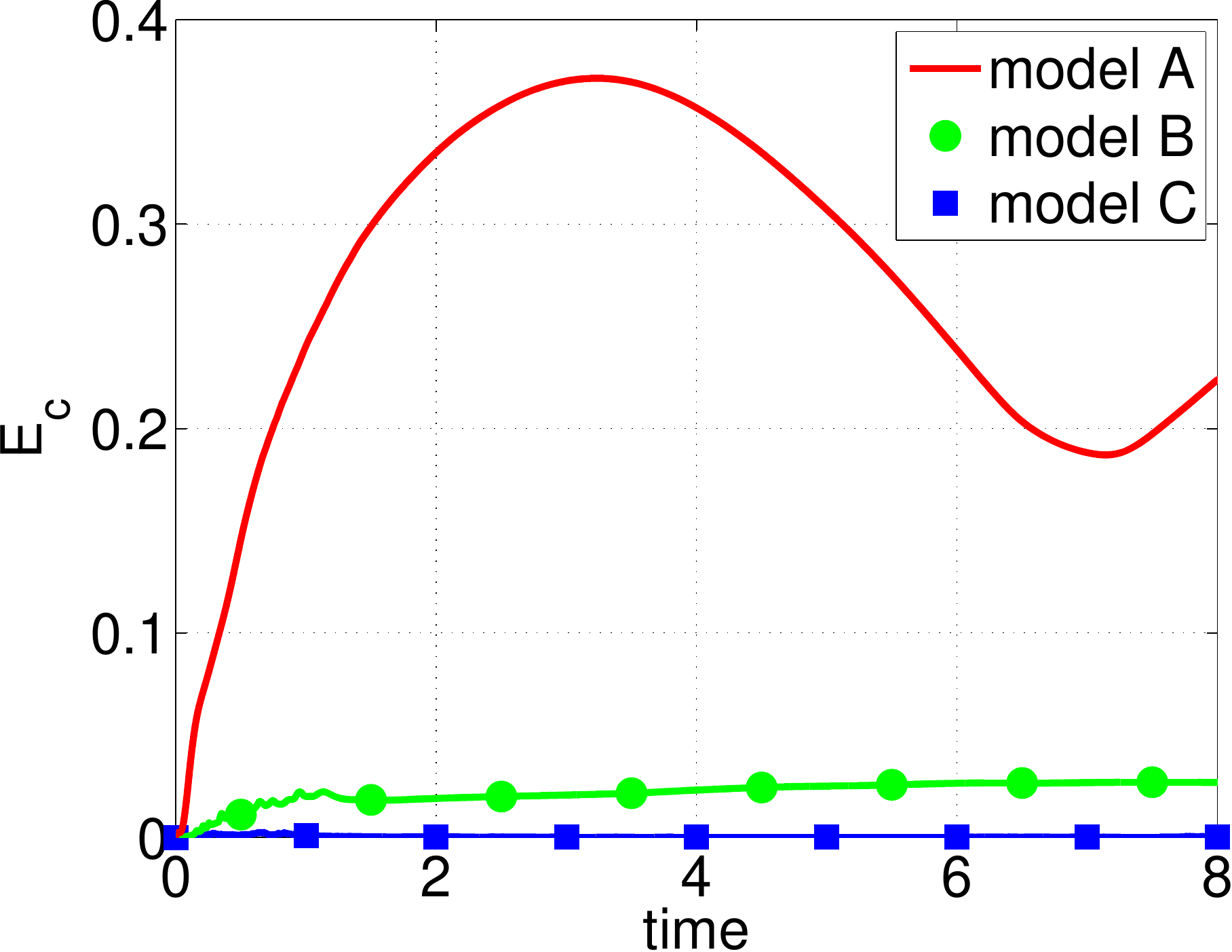}
\includegraphics[width=0.45\textwidth]{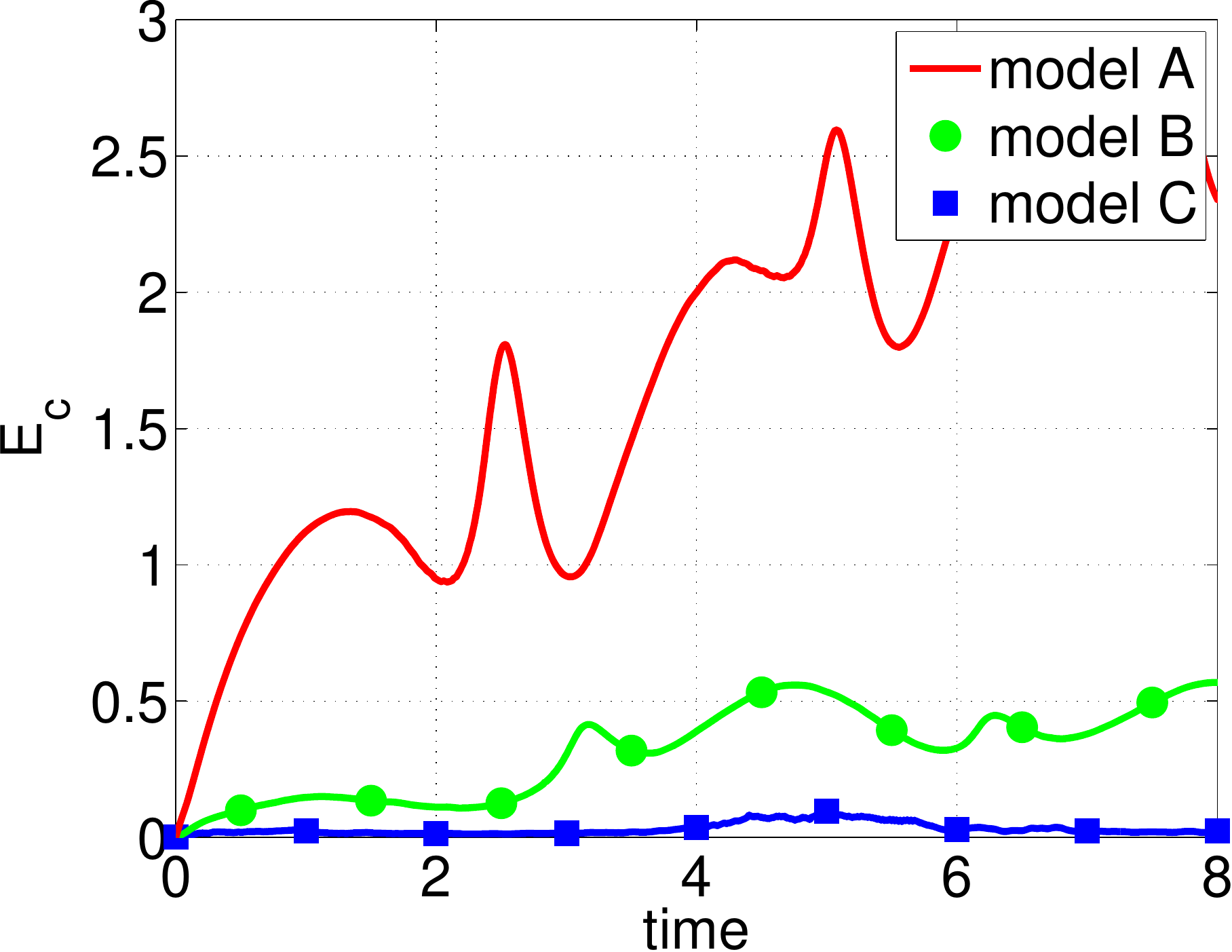}
\caption{Comparison of accumulated stretching $E_c$ for Models A (red), B (green) and C (blue) with $Re=1$ (left) and $Re=1/200$ (right). (color online).
}
\label{fig:model comparison stretching}
\end{figure}

\begin{figure}
\centering
\begin{tabular}{ccc}
model A & model B & model C\\
  \includegraphics[width=0.31\textwidth]{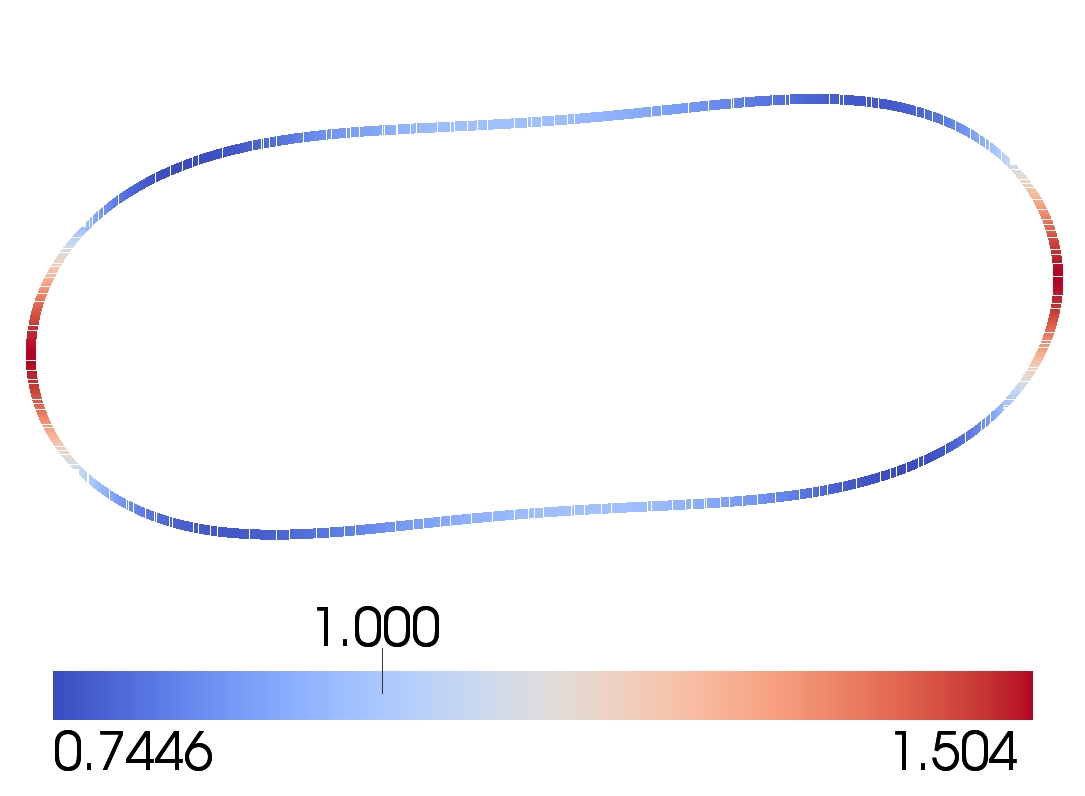} 
&  \includegraphics[width=0.31\textwidth]{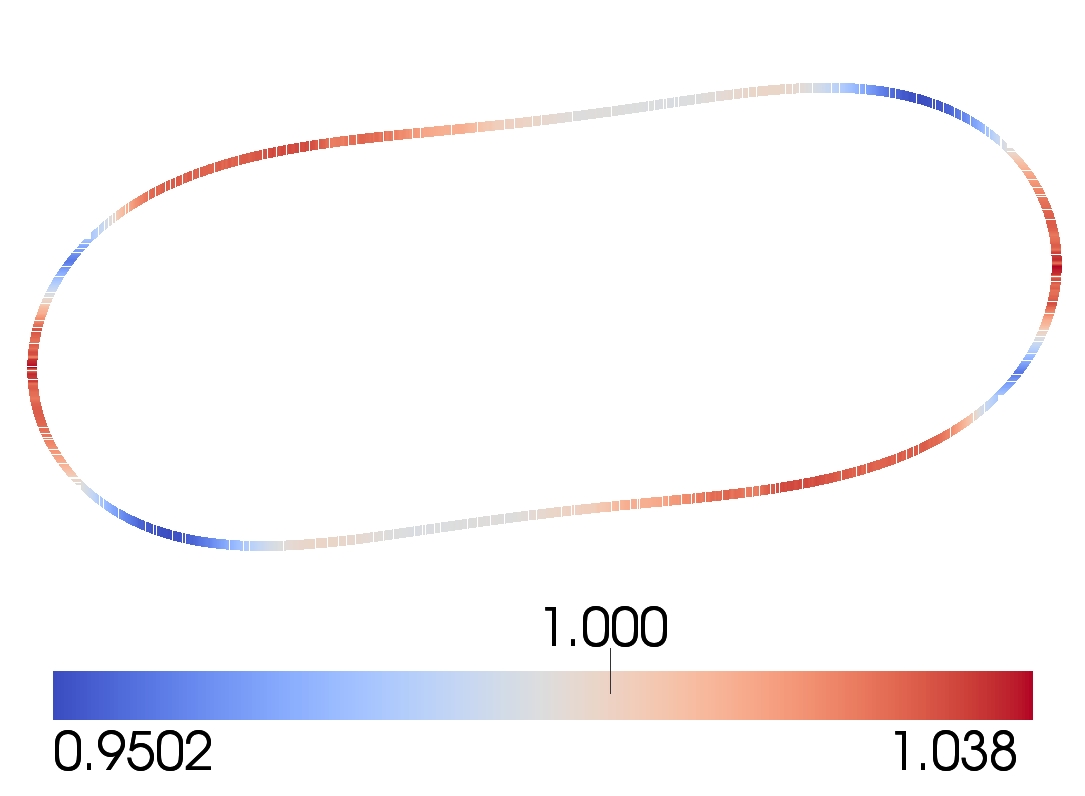}
&  \includegraphics[width=0.31\textwidth]{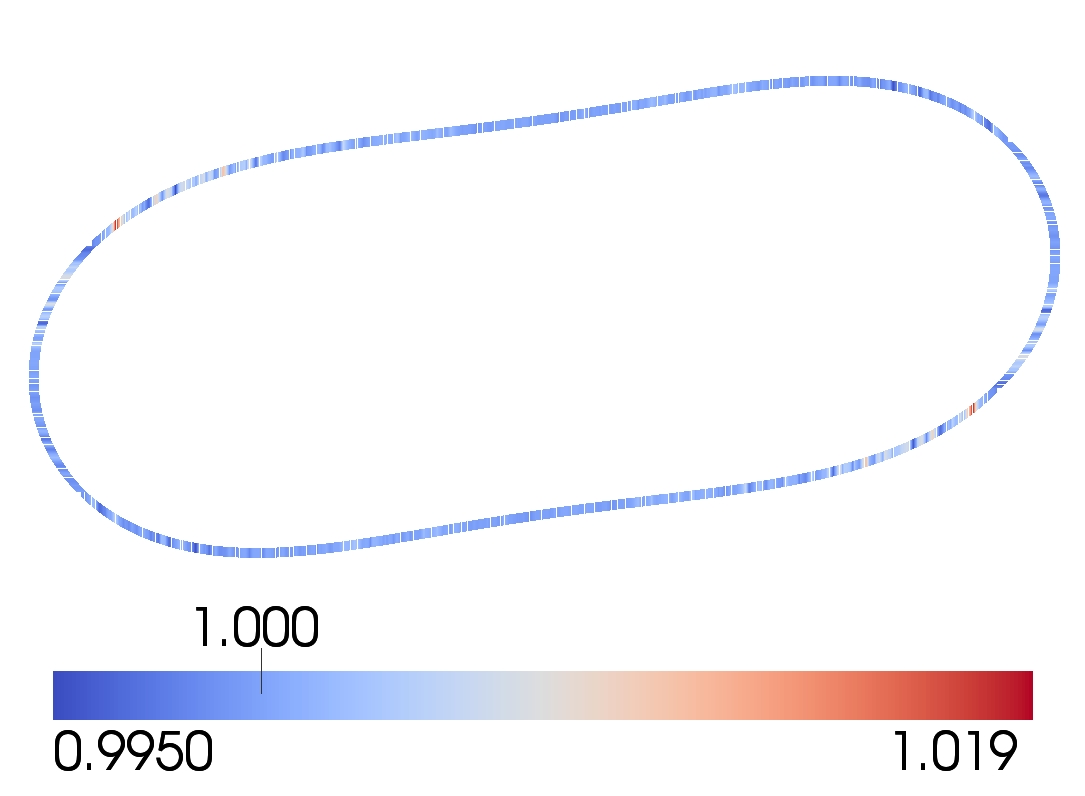} 
\end{tabular}      
  \caption{The value of $c$ along the vesicle interfaces for the different models with $Re=1/200$ at time $t=1$. Note the different scales indicating minimum and maximum value of $c$ as well as the desired value $1.0$. The amount of local stretching/compression decreases from Model A to Model B to Model C. (color online). }
  \label{fig:c}.
\end{figure}

The accumulated stretching $E_c$ for these simulations is depicted in Fig. \ref{fig:model comparison stretching}. As time proceeds, the vesicle interface in Model A is continually stretched.
When $Re=1/200$ (right graph), the amount of stretching is much larger and Model A shows periodic increases in $E_c$, which correspond to tumbling events (e.g., see the inclination angles in Fig. \ref{fig:model comparison angle}). 
The local inextensibility constraint in model B suppresses this significantly, but still the stretching accumulates over time. The local relaxation in Model C effectively controls the accumulation of stretching. Although a small amount of stretching is observed around $t=5$ when the vesicle in Model C tumbles, 
one can how nicely see the stretched vesicle interface is driven back to an unstretched state afterwards. The spatial distributions of $c$ on the interface ($\phi=0$ curve) are shown at time $t=1$ in Fig. \ref{fig:c}. Roughly speaking, in Models A and B the vesicle tips are compressed while the sides are stretched. On the other hand, in Model C the concentration $c\approx 1$ all throughout the vesicle interface. Note that the color scales are different in each case and that the most stretching is observed in Model A as expected.

%
%
%


\section{Conclusions}
\label{conclusions}

We presented a new diffuse interface model for the dynamics of inextensible vesicles in a viscous fluid. Following previous work \cite{Bibenetal_PRE_2003,Beaucourtetal_PRE_2004,Bibenetal_PRE_2005}, we used a local Lagrange multiplier to generate a tension force needed to make the vesicle inextensible. To solve for the local Lagrange multiplier, we introduced a new equation whose solution essentially provides a harmonic extension of the local Lagrange multiplier off the interface while maintaining the local inextensibility constraint near the interface. This is different from the approach taken in  \cite{Bibenetal_PRE_2003,Bibenetal_PRE_2005} where a time-dependent advection-diffusion-reaction equation was used. To make the method more robust we introduced a local relaxation scheme that dynamically corrects stretching/compression errors. In the relaxation scheme, a version of Hooke's law is used where the restoring forces are proportional to the amount of stretching/compression, which is detected by evolving a (surface) concentration field (initialized to one everywhere) and identifying regions where the concentration field deviates from one.  Asymptotic analysis demonstrated that our new system converges to a relaxed version of the inextensible sharp interface system. 

To solve the equations numerically, we developed an efficient algorithm using an operator splitting approach such that the Navier-Stokes equations were implicitly coupled to the diffuse-interface inextensibility constraint. The phase field equations and the local concentration field were solved separately. Spatial discretization was performed using the adaptive finite element toolbox AMDiS \cite{amdis} with the $P^2$/$P^1$ Taylor-Hood element being used for the flow problem, extended by a $P^2$ element for local Lagrange multipliers. $P^2$ elements were also used for the phase field and concentration variables. The resulting nonlinear system was linearized and solved using UMFPACK \cite{umfpack}.

We compared the results from our new model with local inextensibility constraints and relaxation (Model C) to a model without relaxation (Model B) and a previously derived diffuse interface model  \cite{Duetal_DCDS_2007,DuEtAl} that conserved only the total surface area (Model A). Focusing on the dynamics of a single vesicle in shear flow in 2D at two different Reynolds numbers , we demonstrated convergence of the diffuse interface model to the appropriate sharp interface model as the interface thickness tends to zero. We found that the local inextensible constraints generally reduce the amount of rotation the vesicle undergoes so that in the tank-treading regime ($Re=1$), the vesicle from Model A has the smallest inclination angle. Large differences in the dynamics are observed in the tumbling regime ($Re=1/200$) where the local inextensibility constraints in Models B and C delay the tumbling point significantly and increase the length of the tumbling period. The results show that errors in the local inextensibility in Models A and B tend to occur during the fast dynamics of tumbling and accumulate over time. The local relaxation in Model C effectively prevents this accumulation by driving the system back to its equilibrium state when errors in local inextensibility arise.

Future work will use the algorithms presented here to analyze the dependence of the dynamical states of vesicles (tank-treading, tumbling, trembling) on the Reynolds number and other physical parameters (viscosity ratio, density ratio, etc), and the local inextensibility of the interface. We will compare our results with those obtained previously (e.g., \cite{Kelleretal_JFM_1982,Bibenetal_PRE_2003,Beaucourtetal_PRE_2004,Laadharietal_PF_2012,Salacetal_JFM_2012}). We also plan to extend our algorithms to 3D, by replacing the direct UMFPACK solver with a more efficient preconditioned iterative solver for the coupled system, and to incorporate membrane elasticity to provide a more realistic model of RBCs.

\section*{Acknowledgments}

S.A., S.E. and A.V. acknowledge the support of the German Science Foundation within SPP 1506 Al1705/1 and Vo899/11 and
by the European Commission within FP7-PEOPLE-2009-IRSES PHASEFIELD, which J.L. also ackowledges. Further, J.L. is grateful for support from the National Science Foundation Division of Mathematical Sciences and from the National Institutes of Health through grant P50GM76516 for a Center of Excellence in Systems Biology at the University of California, Irvine. Simulations were carried out at ZIH at TU Dresden and JSC at FZ J\"ulich. S.A. and S.E. also thank the hospitality of the Department of Mathematics at the University of California, Irvine where some of this research was conducted.

\section*{Appendix}
\label{Appendix}

Here, we provide justifications for the claims made in Sec. \ref{sec:asymptotics}. In particular, we show that in the inner variables $\mathcal{P}$ has a regular expansion in $\epsilon$, where the leading order term is $\nabla_\Gamma\cdot\mathbf{v}^{(0)}$, and that $\hat{\vct v}^{(0)}_z=\mathbf{0}$.

\paragraph{Regular expansion for $\hat{\mathcal{P}}$} Recall that $\mathcal{P}=\mathbf{P}:\nabla\mathbf{v}$, where $\mathbf{P}=\mathbf{I}-\mathbf{n}\mathbf{n}$ is the tangential projection operator. A straightforward calculation shows that
\begin{equation}
\mathcal{P}=\nabla\cdot\left(\mathbf{P}\mathbf{v}\right)-\left(\nabla\cdot\mathbf{P}\right)\cdot \mathbf{v}
\label{rewrite of mathcalP}
\end{equation}
Therefore, in the inner variables, we obtain
\begin{eqnarray}
\hat{\mathcal{P}}&=&\frac{1}{\epsilon}\mathbf{n}\cdot\left(\mathbf{P}\hat{\vct v}_z\right)+\nabla_\Gamma\left(\mathbf{P}\hat{\vct v}\right)-\left(\nabla_\Gamma\cdot\mathbf{P}\right)\cdot\hat{\vct v},\nonumber\\
&=&\nabla_\Gamma\cdot\left(\mathbf{P}\hat{\vct v}\right)-\left(\nabla_\Gamma\cdot\mathbf{P}\right)\cdot\hat{\vct v},
\label{inner P appendix}
\end{eqnarray}
since $\mathbf{P}_z=0$ and $\mathbf{n}\cdot\left(\mathbf{P}\hat{\vct v}_z\right)=0$. Plugging the inner expansion for $\hat{\vct v}$ into Eq. (\ref{inner P appendix}) we obtain a regular expansion $\hat{\mathcal{P}}=\hat{\mathcal{P}}^{(0)}+\epsilon \hat{\mathcal{P}}^{(1)}+\dots$ and we recognize the first term as
\begin{equation}
\hat{\mathcal{P}}^{(0)}=\nabla_\Gamma\cdot\left(\mathbf{P}\hat{\vct v}^{(0)}\right)+H\hat{\vct v}^{(0)}\cdot\mathbf{n}=\nabla_\Gamma\cdot\hat{\vct v}^{(0)}=\nabla_\Gamma\cdot\mathbf{v}^{(0)}
\label{final}
\end{equation}
as claimed (assuming $\hat{\vct{v}}^{(0)}_z=0$).

\paragraph{Behavior of $\hat{\vct v}^{(0)}$} Writing the incompressibility condition $\nabla\cdot\mathbf{v}=0$  in the inner region, we obtain
\begin{equation}
\frac{1}{\epsilon}\partial_z\left(\hat{\vct v}\cdot\mathbf{n}\right)+\nabla_\Gamma\cdot\hat{\vct v}=0.
\label{incompressibility inner}
\end{equation}
We thus obtain
\begin{eqnarray}
\partial_z\left(\hat{\vct v}^{(0)}\cdot\mathbf{n}\right)&=&0~~~\rm{at}~O(1/\epsilon),\label{incompressible 1}\\
\hat{\vct v}^{(1)}_z\cdot\mathbf{n}+\nabla_\Gamma\cdot\hat{\vct v}^{(0)}&=&0~~~\rm{at}~O(1),\label{incompressible 2}
\end{eqnarray}
and so on. To complete the claim, we need to show that the tangential components of the velocity $\mathbf{P}\hat{\vct v}^{(0)}$ are also independent of $z$. This follows from the viscous term in the Navier-Stokes equations. It can be shown that this term provides the highest order terms in the inner expansion of the Navier-Stokes equations (e.g., see \cite{Duetal_DCDS_2007,DuEtAl}). Thus at the leading order, $O(1/\epsilon^2)$, the Navier-Stokes equations become
\begin{equation}
\partial_{z}\left(\nu\hat{\vct v}_z^{(0)}\cdot\mathbf{n}\right)\mathbf{n}+\partial_{z}\left(\nu\partial_z\left(\mathbf{P}\hat{\vct v}^{(0)}\right)\right)=0.
\label{NS inner}
\end{equation}
Since the first term is zero, we conclude that $\nu\partial_z\left(\mathbf{P}\hat{\vct v}^{(0)}\right)=\rm{constant}$. Taking $z\to\pm\infty$ and using the leading order matching condition (\ref{eq:match1}), we find that the constant is equal to zero, which proves the claim.


\bibliographystyle{plain}
\bibliography{lit}

\end{document}